\documentclass[aps,pra,twocolumn,showpacs,groupedaddress,notitlepage,superscriptaddress,floatfix,nofootinbib,longbibliography]{revtex4-1}
\usepackage{placeins}
\usepackage{graphicx}
\usepackage{subfigure}
\usepackage{amsmath}
\DeclareMathOperator*{\argmax}{argmax}
\usepackage{physics}

\newcommand{\beq}{\begin{equation}}
\newcommand{\eeq}{\end{equation}}
\newcommand{\beqnn}{\begin{equation*}}
\newcommand{\eeqnn}{\end{equation*}}
\newcommand{\bea}{\begin{eqnarray}}
\newcommand{\eea}{\end{eqnarray}}
\newcommand{\ba}{\begin{eqnarray}}
\newcommand{\ea}{\end{eqnarray}}
\newcommand{\beann}{\begin{eqnarray*}}
\newcommand{\eeann}{\end{eqnarray*}}
\newcommand{\bes} {\begin{subequations}}
\newcommand{\ees} {\end{subequations}}

\newtheorem{mytheorem}{Theorem}

\newtheorem{myassumption}{Assumption}

\usepackage{hyperref}
\hypersetup{
    colorlinks=true, 
    linkcolor=cyan,
    citecolor=magenta, 
    filecolor=magenta, 
    urlcolor=cyan,
    runcolor=cyan
}

\begin{document}
\title{Why and when pausing is beneficial in quantum annealing}

\author{Huo Chen}
\affiliation{Department of Electrical Engineering, University of Southern California, Los Angeles, California 90089, USA}
\affiliation{Center for Quantum Information Science \& Technology, University of Southern California, Los Angeles, California 90089, USA}
\author{Daniel A. Lidar}
\affiliation{Department of Electrical Engineering, University of Southern California, Los Angeles, California 90089, USA}
\affiliation{Center for Quantum Information Science \& Technology, University of Southern California, Los Angeles, California 90089, USA}
\affiliation{Department of Chemistry, University of Southern California, Los Angeles, California 90089, USA}
\affiliation{Department of Physics and Astronomy, University of Southern California, Los Angeles, California 90089, USA}

\begin{abstract}
Recent empirical results using quantum annealing hardware have shown that mid anneal pausing has a surprisingly beneficial impact on the probability of finding the ground state for of a variety of problems. A theoretical explanation of this phenomenon has thus far been lacking. Here we provide an analysis of pausing using a master equation framework, and derive conditions for the strategy to result in a success probability enhancement. The conditions, which we identify through numerical simulations and then prove to be sufficient, require that relative to the pause duration the relaxation rate is large and decreasing right after crossing the minimum gap, small and decreasing at the end of the anneal, and is also cumulatively small over this interval, in the sense that the system does not thermally equilibrate. 
This establishes that the observed success probability enhancement can be attributed to incomplete quantum relaxation, i.e., is a form of beneficial non-equilibrium coupling to the environment. 
\end{abstract}

\maketitle

\section{Introduction}
Quantum annealing~\cite{kadowaki_quantum_1998,RevModPhys.80.1061,albash_adiabatic_2018,Hauke:2019aa} stands out among the multitude of concurrent approaches being developed to explore quantum computing, as having achieved the largest scale to date when measured in terms of the sheer number of controllable qubits. Today's commercial quantum annealers feature thousands of superconducting flux qubits and are being used routinely to test whether this approach can provide a quantum advantage over classical computing~\cite{speedup,PhysRevX.6.031015,King:2018aa,Harris:2018aa,Albash:2017aa,Mandra:2017ab,Mott:2017aa}. While there is no consensus that such an advantage has been demonstrated, there is significant progress on the development of ``software" methods that improve quantum annealing performance. Such methods take advantage of the advanced control capabilities of quantum annealers to implement protocols that result in higher success probabilities, shorter times to solution, faster equilibration, etc. Continued progress in this direction is clearly critical as a complementary approach to improving the underlying hardware by reducing physical source of noise and decoherence.

Among the various empirical protocols that have been developed to improve the performance of quantum annealing, such as error suppression and correction~\cite{PAL:13,vinci2015nested,Vinci:2017ab,Pearson:2019aa} and inhomogeneous driving~\cite{Lanting:2017aa,Adame:2018aa,Hsu_2019,Yarkoni:2019aa}, the mid-anneal pausing protocol stands out as particularly powerful. Pausing superficially resembles the idea of slowing down near the minimum gap, as in the optimal schedule for the Grover problem~\cite{Roland:2002ul,albash_adiabatic_2018}, but the context here is entirely different due to the fact that pausing happens in an open system subject to thermal relaxation. The first study~\cite{marshall_power_2019} to systematically test this approach empirically using a 
D-Wave 2000Q device~\cite{DW2KQ}, demonstrated a dramatic improvement in the probability of finding the ground state (i.e., the success probability) when an anneal pause was inserted shortly after crossing the minimum gap. Follow-up studies confirmed that pausing is advantageous on different problems such as portfolio optimization problems~\cite{venturelli_reverse_2019} and training deep generative machine learning models~\cite{vinci2019path}. Numerical studies \cite{passarelli_improving_2019,passarelli_reverse_2020} of the $p$-spin model also agree with these empirical results. However, despite a useful qualitative explanation offered for the thermalization mechanism by which pausing improves success probabilities~\cite{marshall_power_2019}, a thorough analysis of the exact mechanism of this phenomenon is still lacking. Here we provide such an analysis, and identify sufficient conditions for pausing to provide an enhancement.

Our analysis is based on a detailed investigation of a quantum two-level system model coupled to an Ohmic bath. The two-level system can be either a single qubit or a multi-qubit system whose lowest two energy levels are separated by a large gap from the rest of the spectrum. The analysis builds on tools from the theory of open quantum systems~\cite{Breuer:book,alicki_quantum_2007,Lidar:2019aa}, specifically master equations appropriate for time-dependent (driven) Hamiltonians~\cite{albash_quantum_2012,albash_decoherence_2015,smirnov_theory_2018,Dann:2018aa,mozgunov_completely_2019,nathan2020universal}. Through numerical investigation we identify a set of sufficient conditions, stated in term of the properties of the relaxation rate along the anneal, and prove a theorem guaranteeing that it is advantageous to pause mid-anneal. The advantage gained is a higher success probability than is attainable without pausing. 

We thus establish, in a rigorous sense, that there exists a non-trivial optimal pausing point under a set of reasonable assumptions. We do not identify the optimal pausing duration, but we do prove that the optimal pausing point occurs after the minimum gap, in accordance with the prior empirical and numerical evidence. This result is stated in Theorem~\ref{thm:opt-pausing} below, which can be summarized as saying that an optimal pausing point exists if the relaxation rate right after the minimum gap is large relative to the pause duration but small at the end of the anneal, decreases right after crossing the minimum gap and also at the end of the anneal, and is also cumulatively small over this interval, in the sense that the system does not fully thermally equilibrate.

The structure of this paper is as follows. In Sec.~\ref{sec:1qmodel} we define our model of the two-level system. In the multi-qubit case this involves deriving an effective Hamiltonian for the projection to the low energy subspace of the full Hamiltonian. We introduce a certain parametrization of the gap and the geometric phase that ensures the problem is hard for quantum annealing, in the sense that the success probability is low even on a timescale that is large compared to the inverse of the minimum spectral gap along the anneal. In Sec.~\ref{sec:open_system} we treat the same model as an open quantum system using master equation techniques, specifically the Redfield equation with and without the rotating wave approximation, and the adiabatic master equation. Then, in Sec.~\ref{sec:pausing} we introduce a pause into the annealing schedule and study its effects. We first demonstrate numerically that an optimal pausing position exists before the end of of the anneal, depending on the monotonicity properties of the relaxation rate after the minimum gap is crossed. Building on these observations we then prove a theorem establishing sufficient conditions for the existence of such an optimal pausing point. We conclude in Sec.~\ref{sec:conc}, and present additional technical details in the Appendix.

\section{``Hard" single-qubit and multi-qubit closed system models}
\label{sec:1qmodel}

In this section we consider two scenarios: single qubit annealing, and a projected two-level system (TLS) arising from multi-qubit annealing. We define a model that makes these problems ``hard" for quantum annealing, in the sense of a small success probability even over a timescale that is long compared to the heuristic adiabatic timescale (given by the inverse of the minimum gap along the anneal path).

\subsection{Single qubit}
\label{sec:single_qubit_model}

We write the single qubit annealing Hamiltonian in the form
\begin{equation}
H_{\mathrm{S}}(s) = -\frac{1}{2} \big[ A(s)Z +  B(s)X \big] \ ,
\label{eq:single_HS}
\end{equation}
where $s=t/t_f$ is the dimensionless instantaneous time, $t$ is the actual time, $t_f$ is the total anneal time, and $A(s)$ and $B(s)$ are the annealing schedules. Note that we permuted the Pauli matrices $X$ and $Z$ of the conventional single qubit annealing Hamiltonian in order to have the same expression for both the single qubit and projected TLS cases. After transforming to the adiabatic frame~\cite{munoz-bauza_double-slit_2019}, the dimensionless interaction picture Hamiltonian becomes:

\begin{equation}
\label{eq:single_qubit_adiabatic_hamiltonian}
    \tilde{H}_\mathrm{S}\pqty{s} = \frac{1}{2}\pqty{\dv{\theta}{s}Y-t_f\Omega\pqty{s}Z}\ ,
\end{equation}
where $\Omega(s)$ and $\theta(s)$ are a reparameterization of the annealing schedules: $A(s) = \Omega(s)\cos\theta(s)$ and $B(s) = \Omega(s)\sin\theta(s)$.
(see Appendix~\ref{app:single_qubit_adiabatic_frame} for a detailed explanation). We call $\theta(s)$ the annealing angle and $\dot{\theta}(s)$ the angular progression. The term $\dot{\theta}Y/2$ has its origin as a geometric phase~\cite{vinci_non-stoquastic_2017}. Loosely, $\Omega(s)$ corresponds to the time-dependent gap and $\mathrm{d}\theta/\mathrm{d}s$ corresponds to how fast/slow the Hamiltonian changes. For a typical single qubit annealing process, the boundary condition $\theta(0)=0$ and $\theta(1)=\pi / 2$ needs to be satisfied (noticing that in Eq.~\eqref{eq:single_HS} we permuted the Pauli $X$ and $Z$ matrices in the standard notation).

\subsection{Projected TLS from a multi-qubit model}
\label{sec:multi_qubit_adiabatic_frame}

For general multi-qubit annealing, the Hamiltonian is
\begin{equation}
\label{eq:multi_HS}
    H_\mathrm{S}(s) = A(s)H_{\mathrm{d}} + B(s)H_{\mathrm{p}} = \sum_n E_n\pqty{s}\dyad{n(s)} \ .
\end{equation}
where $\Bqty{\ket{n\pqty{s}}}$ is the instantaneous energy eigenbasis and $E_n\pqty{s}$ are the instantaneous energies. $H_\mathrm{d}$ and $H_\mathrm{p}$ are the driver and problem Hamiltonian, respectively. Henceforth we assume that $H_\mathrm{S}=(H_\mathrm{S})^T$, i.e., that $H_\mathrm{S}(s)$ is real for all $s$. The system density matrix can be written in the instantaneous energy eigenbasis:
\begin{equation}
\label{eq:density_matrix_energy_eigen}
    \rho\pqty{s} = \sum_{nm}\rho_{nm}\dyad{n}{m} \ .
\end{equation}
We call the associated matrix $\tilde{\rho} = \bqty{\rho_{nm}}$ the density matrix in the adiabatic frame, and show in Appendix~\ref{app:multi_qubit_adiabatic_frame} that it obeys the von Neumann equation $\dot{\tilde{\rho}} = -i\comm{\tilde{H}}{\tilde{\rho}}$ with the effective Hamiltonian
\begin{equation}
\label{eq:effective_H}
    \tilde{H} = \begin{pmatrix}
    t_fE_0 & -i \braket{0}{\dot{1}} & \dots\\
    i\braket{0}{\dot{1}} & t_fE_1 & \dots\\
    \vdots& & \ddots
    \end{pmatrix}
\end{equation}
If we truncate the effective Hamiltonian~\eqref{eq:effective_H} to the lowest two energy levels and shift it by a constant term, we find:
\begin{equation}\label{eq:effective_H_pauli}
    \tilde{H}_2 = \braket{0}{\dot{1}} Y  - \frac{t_f\Omega\pqty{s}}{2}Z ,
\end{equation}
where $\Omega(s)=E_1-E_0$ is the energy gap between the lowest two energy levels. We call this the projected TLS Hamiltonian. An alternative way to derive this effective Hamiltonian is via the well-known adiabatic intertwiner (see, e.g., Ref.~\cite{rezakhani_intrinsic_2010}). This TLS approximation is valid when (i) there is a large gap separating the two-level subspace from higher excited state (where ``large" is in the sense of the adiabatic theorem~\cite{Jansen:07}), and (ii) the geometric terms connecting the two-level subspace to the higher levels in the adiabatic frame are negligible~\cite{Vinci:2017aa}. One may also invoke the Schrieffer-Wolff transformation to establish similar conditions~\cite{Bravyi20112793,Consani:2020aa}.

Since the annealing Hamiltonian [Eq.~\eqref{eq:multi_HS}] is real, $\braket{0}{\dot{1}}$ in Eq.~\eqref{eq:effective_H} is also real. This effective Hamiltonian is equivalent to Eq.~\eqref{eq:single_qubit_adiabatic_hamiltonian} with $\braket{0}{\dot{1}}$ playing the role of $\dot{\theta}/2$. Thus, we can define the angular progression as $\dot{\theta} = 2\braket{0}{\dot{1}}$ for the projected TLS.  Having done so, a general TLS Hamiltonian can also be written as Eq.~\eqref{eq:single_qubit_adiabatic_hamiltonian}, where the annealing angle $\theta$ in the general case does not need to satisfy the same boundary condition as in the single qubit case.

\subsection{Hard Problem Instances from Gap and Angular Progression Considerations}
\label{sec:hard}

We call an instance ``easy" when a high ground state probability is achieved within an annealing time that is much shorter than the timescale set by the inverse of the minimum gap along the anneal (we refer to this as the ``heuristic" adiabatic condition; it is not to be confused with the rigorous adiabatic condition, which provides a sufficient condition for convergence to the ground state~\cite{Jansen:07,lidar_adiabatic_2009}). Conversely, we call an instance ``hard" when the ground state probability is low for such an annealing time. 

Our strategy is to create toy models that share the same features as certain known hard examples for quantum annealing~\cite{dickson_thermally_2013, passarelli_improving_2019}. By closely examining those problems, we identify one crucial characteristic they share: a sharp peak in the angular progression appears at the minimum gap, along with a $\pi$ jump of the annealing angle $\theta$ (see Appendix~\ref{app:multi_qubits_example} where these examples are illustrated).

We take a reverse engineering approach by first specifying the analytic form of the gap $\Omega(s)$ and angular progression $\dot{\theta}\pqty{s}$ in the adiabatic frame. The gap is parametrized as a Gaussian in the form 
\begin{equation}
\label{eq:gaussian_gap}
    \Omega(s)= E_0 \pqty{1-\pqty{1-\Delta}e^{-\frac{(s-\mu_g)^2}{2\alpha^2_g}}} \ ,
\end{equation}
where the parameters $\Delta$, $\mu_g$ and $\alpha_g$ respectively control the gap size, position and width. The angular progression can also be chosen as Gaussian
\begin{equation}
\label{eq:gaussian_angular_momentum}
    \dot{\theta}(s) = Ce^{-\frac{(s-\mu_\theta)^2}{2\alpha_\theta^2}}
\end{equation}
with position and width parameter $\mu_\theta$ , $\alpha_\theta$. The normalization constant $C$ is chosen according to the boundary condition. We will discuss both the single qubit and projected TLS cases.

\subsection{Single qubit}

\begin{figure}[ht]
\includegraphics[width=\columnwidth]{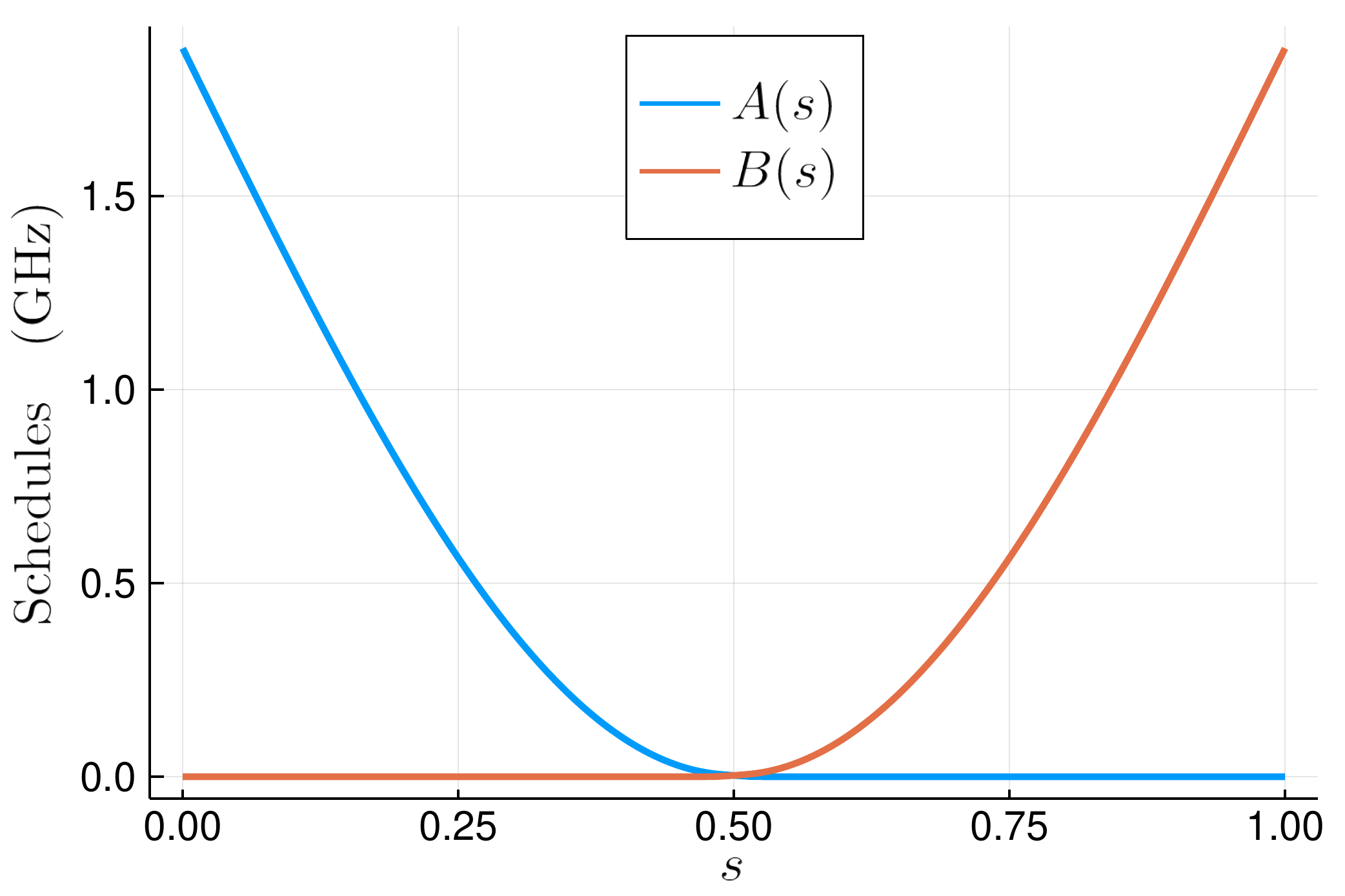}
\caption{Example schedules with the following parameter choices: $\mu_\theta=\mu_g=0.5$, $\alpha_g = 0.5$, $E_0=15/\pi$ GHz, $\Delta=0.001$ and $\alpha_\theta = 1/100$. 
}
\label{fig:schedules}
\end{figure}

As a consequence of the aforementioned boundary conditions $\theta(0)=0$ and $\theta(1)=\pi / 2$. the normalization constant is
\begin{equation}
    C = \sqrt{\frac{\pi}{2}}\frac{1}{2\alpha_\theta} \ ,
\end{equation}
and the annealing angle resulting from Eq.~\eqref{eq:gaussian_angular_momentum} is
\begin{equation}
    \theta(s) = \frac{\pi}{4}\bigg[\erf\left(\frac{\mu_\theta}{2\alpha_\theta}\right)+\erf\left(\frac{s-\mu_\theta}{\sqrt{2}\alpha_\theta}\right)\bigg] \ .
\end{equation}
In order to ensure the hardness of the problem and monotonic schedules, we need to overlap the peak region of $\dot{\theta}$ and the minimum gap, i.e., $\mu_\theta \approx \mu_g$. In such a region, the diabatic term is much larger than the adiabatic term and the Landau-Zener transition is strong. A choice of such schedules is illustrated in Fig.~\ref{fig:schedules}. 
It is important to note that, in this construction, the hardness of the problem is not solely determined by the minimum gap $\Delta$. Indeed, the rigorous adiabatic condition involves the derivative of the Hamiltonian as well~\cite{Jansen:07,lidar_adiabatic_2009}. An example of an easy instance with a small minimum gap is given in Appendix~\ref{app:const_dtheta}. 

\FloatBarrier

\subsection{Projected TLS}
The boundary conditions for the projected TLS are different from the single qubit case because there is no simple relation between the schedules and the annealing angle. However, a common feature of the small gap instances~\cite{dickson_thermally_2013, passarelli_improving_2019} is a localized pulse of angular progression that is present at the minimum gap. Also, this pulse induces a step-function like $\pi$ shift of the annealing angle across this region, leading to a near-perfect Landau-Zener transition. The simplest toy model we can construct is to keep the Gaussian form of the gap [Eq.~\eqref{eq:gaussian_gap}] and angular progression [Eq.~\eqref{eq:gaussian_angular_momentum}] but use a different boundary condition $\theta(0)=0$ and $\theta(1)=\pi$, which comes from the examination of both the $p$-spin model~\cite{passarelli_improving_2019} and the D-Wave $16$-qubit gadget problem~\cite{dickson_thermally_2013}. In this case, the normalization constant becomes $C=\sqrt{\pi}/\sqrt{2}\alpha_\theta$.

However, we stress that the core of this construction is the  angular progression pulse at the minimum gap. There is no constraint on $\dot{\theta}\pqty{s}$ at other $s$ as long as the system can follow its eigenstates before and after the pulse. In fact, in the problem studied in Ref.~\cite{dickson_thermally_2013}, the annealing angle first gradually decreases to a non-zero value before the $\pi$ jump  (see Appendix \ref{app:multi_qubits_example}).

\section{Open System Model}
\label{sec:open_system}

For the open quantum system model, we directly start with the multi-qubit case. We adopt a standard noise model for quantum annealing: each qubit couples to a bosonic bath via a system operator $O$:
\begin{equation}
\label{eq:interaction_term}
    H_{\mathrm{SB}} = \sum_{\alpha k} g_{\alpha k}O_\alpha \otimes (b_{\alpha k}^\dagger+b_{\alpha k}) = \sum_\alpha g_\alpha O_\alpha \otimes \mathcal{B}_\alpha \ .
\end{equation}
Here $g_\alpha$ and $O_\alpha$ are dimensionless and $\mathcal{B}_\alpha$ has dimensions of  energy. The parameters $g_\alpha$ serve as expansion variables, which can later be set to one. After moving to the adiabatic frame, the system-bath interaction becomes
\begin{equation}
    \tilde{H}_{\mathrm{SB}}\pqty{s} = t_f\sum_{\alpha m n} g_\alpha O_{\alpha}^{mn}(s)\otimes\mathcal{B}_\alpha\ ,
\end{equation}
where
\begin{equation}
    O_\alpha^{mn}(s) = \mel{m(s)}{O_\alpha}{n(s)} \dyad{m}{n}\ .
\end{equation}
By defining
\begin{equation}
    S_\alpha(s) = \sum_{mn} O_{\alpha}^{mn}(s) \ ,
\end{equation}
the total projected TLS Hamiltonian in the adiabatic frame can be further simplified as
\begin{equation}\label{eq:open_system_hamiltonian}
    \tilde{H} = \frac{1}{2}\pqty{\dot{\theta}\pqty{s}Y-t_f\Omega\pqty{s}Z} + \sum_\alpha g_\alpha t_f S_\alpha\pqty{s}\otimes\mathcal{B}_\alpha + H_\mathrm{B} \ .
\end{equation}
From now on, for conciseness we will omit the the tilde symbol for adiabatic frame operators. We investigate three approaches for solving the open system dynamics.

\subsection{Redfield Equation}
\label{sec:redfield}

Before proceeding, we define the bath correlation function
\begin{equation}
    C_{\alpha\alpha'}(s,s') = \Tr[\mathcal{B}_\alpha(s)\mathcal{B}_{\alpha'}\pqty{s'}\rho_{\mathrm{B}}] = C_\alpha^*(s',s)
\end{equation}
in terms of the rotated $\mathcal{B}_\alpha$ operator
\begin{equation}
    \mathcal{B}_\alpha(s) = U^\dagger_{\mathrm{B}}(s) \mathcal{B}_\alpha U_\mathrm{B}(s) \ ,
\end{equation}
where $U_\mathrm{B}(s) = \exp[-it_fH_\mathrm{B}s]$ is the free bath evolution. 
We call a set of $\Bqty{\mathcal{B}_\alpha(s)}_{\alpha}$ independent if $C_{\alpha\alpha'}(s,s') = \delta_{\alpha\alpha'} C_{\alpha}(s, s')$ $\forall \alpha, \alpha'$, and  identical if $C_{\alpha}(s, s') = C(s, s')$ $\forall \alpha$.

By assuming $\Bqty{\mathcal{B}_\alpha(s)}_{\alpha}$ are independent, the Redfield equation in the adiabatic frame can be shown to be~\cite{munoz-bauza_double-slit_2019}:
\begin{align}
	\label{eq:adiabatic_tcl2}
	\dot{\rho}_S(s) &= -i\comm{H_\mathrm{S}(s)}{{\rho}_S(s)}  \\
	&\qquad -\sum_\alpha(g_\alpha t_f)^2 \comm{S_\alpha\pqty{s}}{\Lambda_{\alpha}(s)\rho_S(s)} + \textrm{h.c.}\ , \notag
\end{align}
where 
\begin{equation}
\label{eq:TCL_lambda}
\Lambda_\alpha(s) = \int_0^s \dd{s'} C_\alpha(s,s')U(s,s')S_\alpha\pqty{s'}U^\dagger(s,s') \ ,
\end{equation}
and
\begin{equation}
    U(s,s') = T_+\exp{-it_f\int_{s'}^s H_\mathrm{S}(s'') \dd{s''}} \ ,
\end{equation}
and $T_+$ denotes time-ordering.
An important observation is that, after moving to the adiabatic frame, the transformed system Hamiltonian 
[Eq.~\eqref{eq:single_qubit_adiabatic_hamiltonian}] 
has a different gap than the original one [Eq.~\eqref{eq:single_HS}], due to the rescaling by $t_f$. We define the new gap (in energy units) in the adiabatic frame as
\begin{equation}
\label{eq:virtual_gap}
    \Delta\pqty{s} = \sqrt{\dot{\theta}^2\pqty{s}/t_f^2+\Omega^2\pqty{s}} \ .
\end{equation}

\subsection{Redfield Equation with Rotating Wave Approximation}
\label{sec:redfieldRWA}

From our construction of $\Omega\pqty{s}$ and $\dot{\theta}\pqty{s}$ [Eqs.~\eqref{eq:gaussian_gap} and \eqref{eq:gaussian_angular_momentum}], it follows that at the minimum gap point of $s=\alpha_g$, $\Delta\pqty{s}$ is large. As a consequence, we can safely apply the rotating wave approximation (RWA) with the adiabatic frame Redfield Eq.~\eqref{eq:adiabatic_tcl2} without worrying about the presence of a small gap. After the RWA, Eq.~\eqref{eq:TCL_lambda} becomes
\begin{align}
		\dot{\rho}_S &= -i\comm{H_\mathrm{S}\pqty{s}+H_{\mathrm{LS}}}{{\rho}_S} \notag \\
		\label{eq:TCL2_RWA}
		&\qquad -\Gamma_{\mathrm d} \big(\rho_{ba}\dyad{b}{a}+\rho_{ab}\dyad{a}{b}\big)  \\
		&\qquad +\Gamma_{\mathrm t} \pqty{\rho_{aa}-e^{-\beta\Delta}\rho_{bb}}\big(\dyad{b}{b}-\dyad{a}{a}\big) \ , \notag
\end{align}
where $\rho_{ab}=\matrixel{a}{\rho_S}{b}$ with $\{\ket{a},\ket{b}\}$ being the ground and excited states of $H_\mathrm{S}\pqty{s}$. All quantities in Eq.~\eqref{eq:TCL2_RWA} are $s$-dependent, and the effective dephasing and thermalization rates $\Gamma_{\mathrm d}$ and $\Gamma_{\mathrm t}$, respectively, are given by~\cite{munoz-bauza_double-slit_2019}:\footnote{The expression we arrive at here for $\Gamma$ here is slightly different from Ref.~\cite{munoz-bauza_double-slit_2019} since here the RWA is done in the adiabatic frame, while in Ref.~\cite{munoz-bauza_double-slit_2019} the RWA is done in an additional rotating frame.}
\bes
\begin{align}
\label{eq:Gamma_d}
    \Gamma_{\mathrm d}\pqty{s} &= \frac{t_f}{2}\Gamma_{\mathrm t}\pqty{s}\big(1+e^{-\beta \Delta(s)}\big) \notag\\ &+\frac{t_f}{2}\sum_\alpha\gamma_\alpha\pqty{0}\pqty{S_\alpha^{aa}-S_\alpha^{bb}}^2\\
    \Gamma_{\mathrm t}\pqty{s} &= t_f\sum_{\alpha} g_{\alpha}^2 \gamma_\alpha\pqty{\Delta}\abs{S_\alpha^{ab}}^2
\end{align}
\ees
where the projected system-bath coupling operators are:
\begin{equation}
\label{eq:proj-SB-ops}
    S_\alpha^{ab} = \mel{a}{S_\alpha\pqty{s}}{b} \ .
\end{equation}
The Lamb shift is:
\begin{equation}
	H_{\mathrm{LS}}(s) = \sum_\alpha g^2_\alpha t_f(\mathcal{S}_\alpha\pqty{\Delta(s)}\dyad{b}+\mathcal{S}_\alpha\pqty{-\Delta(s)}\dyad{a})\ .
	\label{eq:H_LS}
\end{equation}
The functions $\gamma_\alpha\pqty{\omega}/2$ and $\mathcal{S}_\alpha\pqty{\omega}$ are the real and imaginary parts of the noise spectral density (the one-sided Fourier transform of the bath correlation function). 

\subsection{Adiabatic Master Equation}
Outside the peak region of the angular progression, Eq.~\eqref{eq:TCL2_RWA} becomes the adiabatic master equation (AME)~\cite{albash_quantum_2012}.
The AME is a special case of Eq.~\eqref{eq:TCL2_RWA}. It can be derived by ignoring the geometric part of the Hamiltonian in the unitary part of the Redfield Eq.~\eqref{eq:adiabatic_tcl2}, which holds in the adiabatic limit $t_f \gg 1$. It has the same form as Eq.~\eqref{eq:TCL2_RWA}, with $\Delta\pqty{s}$ being the physical gap [Eq.~\eqref{eq:virtual_gap} with $\dot\theta=0$] and $\Bqty{\ket{a}, \ket{b}}$ being the instantaneous eigenstates.

\section{Pausing}
\label{sec:pausing}
So far we only considered the case of a linear annealing parameter $s=t/t_f$. In this section we study the effect of including a pause.

\subsection{Model}
\label{sec:model}

\begin{figure}[t]
\includegraphics[width=\columnwidth]{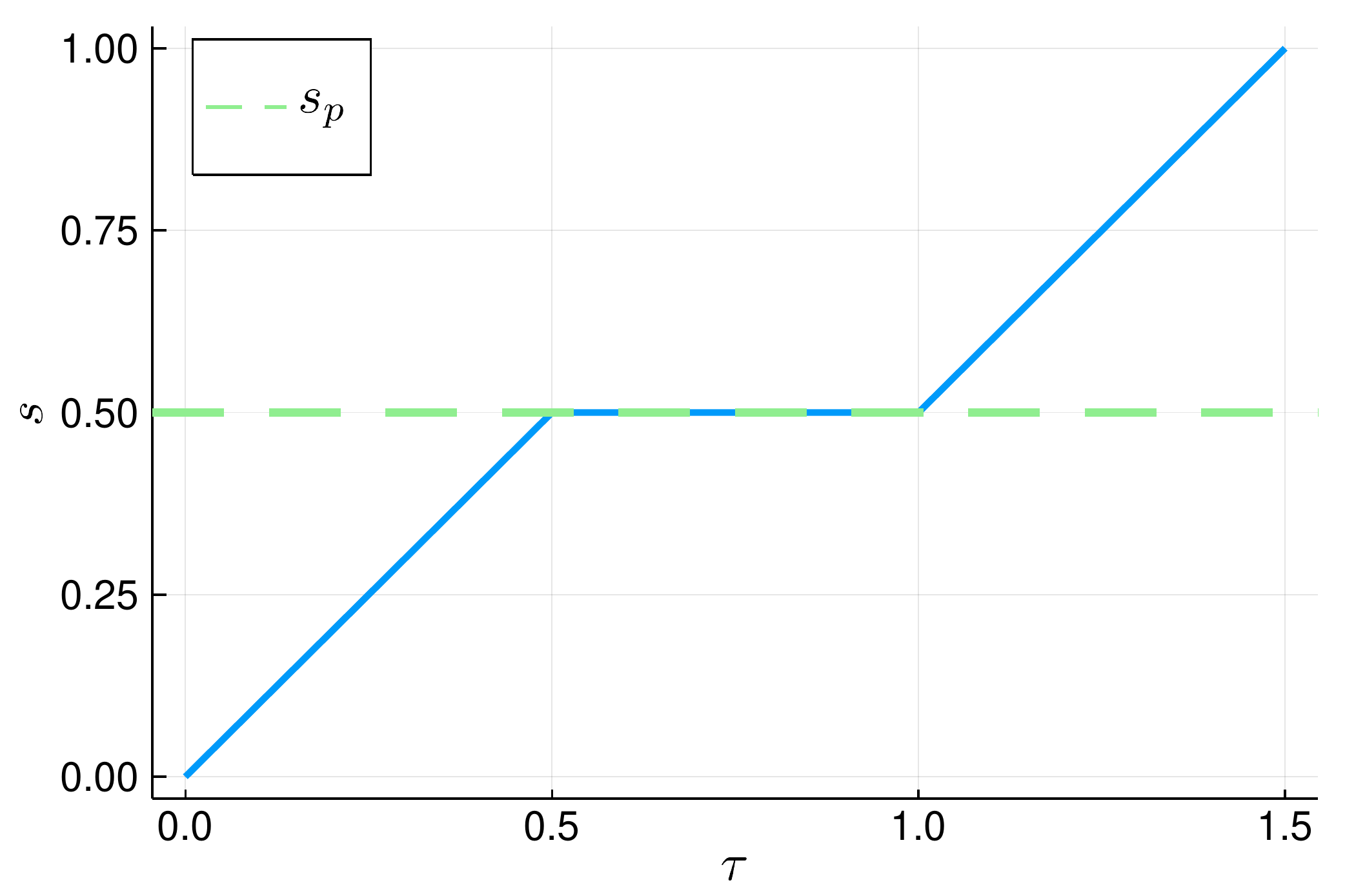}
\caption{ Example of annealing parameter $s$ against dimensionless time $\tau = t/t_f$. The pause position $s_p$ and duration $s_d$ are choose as $s_p=0.5$ and $s_d=0.5$. 
}
\label{fig:pausing}
\end{figure}

To incorporate a pause, let us define $s = s(\tau;s_p,s_d)$ where $\tau = t / t_f$ is the dimensionless time, and where $s_p$ and $s_d$ are the pausing position and pausing duration, respectively. The explicit form of $s(\tau;s_p,s_d)$ is given below, where from now on we suppress the explicit dependence on $s_p$ and $s_d$ for simplicity, and is illustrated in Fig.~\ref{fig:pausing}:

\begin{equation}
    s(\tau) = \begin{cases}
    \tau \quad &\tau \leq s_p \\
    s_p \quad &s_p < \tau \leq s_p + s_d \\
    \tau - s_d \quad &s_p + s_d < \tau \leq 1+s_d
    \end{cases} .
    \label{eq:cases}
\end{equation}
Note that pausing increases the total annealing time to 
\beq
t'_f = \tau_f t_f \ , \qquad 
\tau_f = 1 + s_d\ .
\label{eq:tau_f}
\eeq 
To prevent confusion, \emph{henceforth we will denote by $Q(s)$ the original quantity and by $Q(\tau)$ the corresponding paused quantity}, where $Q$ can be any function or operator. 

The dimensionless Hamiltonian in the adiabatic frame then becomes:
\begin{equation}\label{eq:pause_H}
    H\pqty{\tau} = \frac{1}{2}\pqty{\dv{\theta}{\tau}X- t'_f \Omega(\tau) Z} +  t'_f \sum_\alpha g_\alpha S_\alpha\pqty{\tau}\otimes\mathcal{B}_\alpha + H_\mathrm{B} \ ,
\end{equation}
where $\mathrm{d}\theta /\mathrm{d}\tau$ can be calculated using the chain rule:
\begin{equation}
    \dv{\theta}{\tau} = \begin{cases}
        0 \quad& s_p<\tau\leq s_p+s_d\\
        \mathrm{d}\theta/\mathrm{d}s \quad& \mathrm{elsewhere} 
    \end{cases} \ .
\end{equation}

\subsection{Numerical Results}
\label{sec:numerical}
\begin{figure}[t]
    \includegraphics[width=\columnwidth]{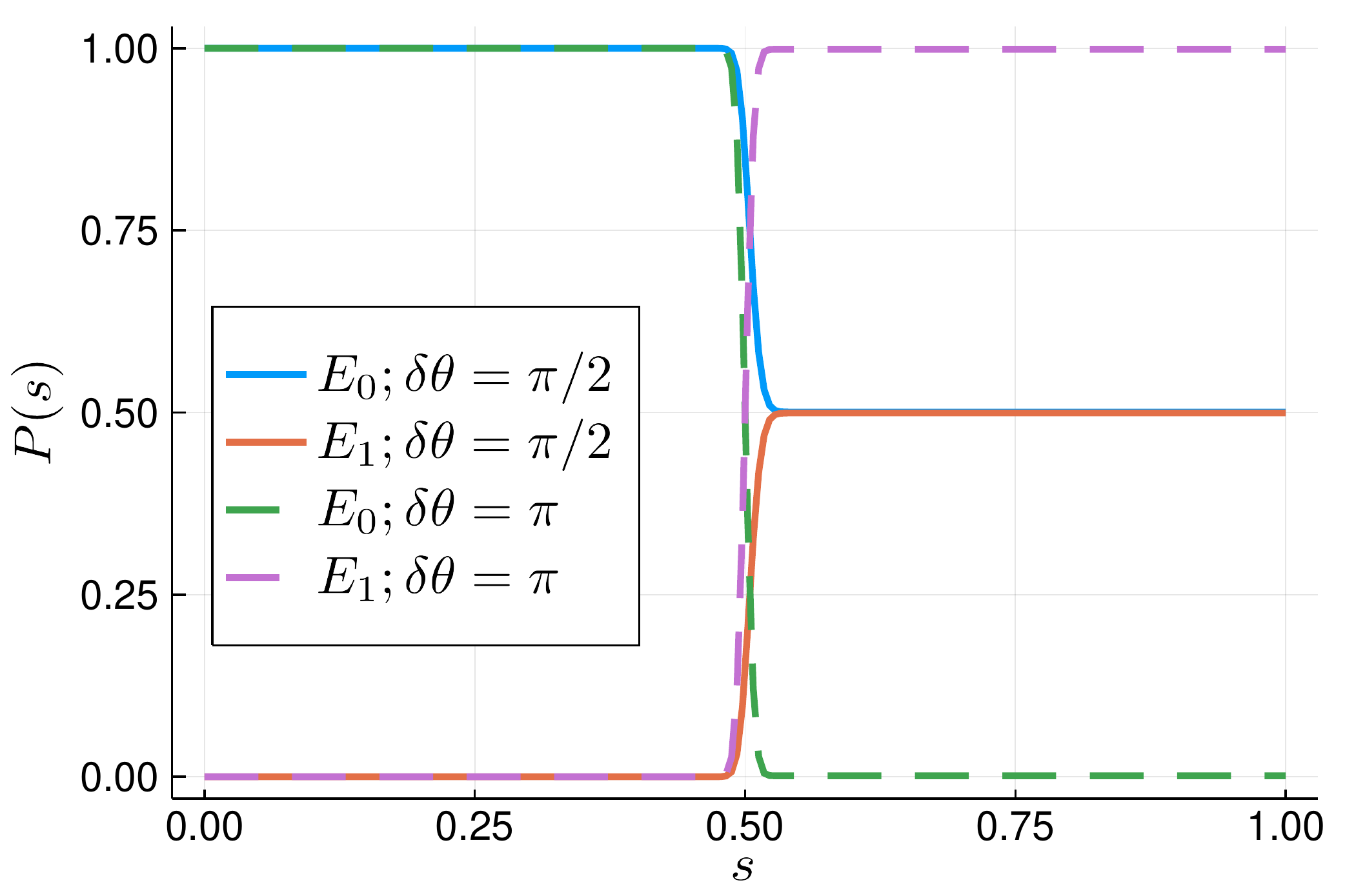}
    \caption{Populations of instantaneous eigenstates during an anneal with total time $t_f = 100 (\mathrm{ns})$. $\delta \theta$ denotes the jump of the annealing angle $\theta$ across the minimum gap region. In our model, the magnitude of this jump is directly determined by the boundary condition.
    }
    \label{fig:close_inst}
\end{figure}

For our numerical simulations, we use the Gaussian gap [Eq.~\eqref{eq:gaussian_gap}] and angular progression [Eq.~\eqref{eq:gaussian_angular_momentum}] with two different boundary conditions: $\theta\pqty{1} = \pi/2$ and $\theta\pqty{1}=\pi$. The other parameters are the same as in Fig.~\ref{fig:schedules}. The instantaneous populations during a 100(ns) anneal are shown in Fig.~\ref{fig:close_inst}. We observe that the role of boundary conditions in our setup (with a single minimum gap) is to determine the portion of population transferred to the excited state when traversing the minimum gap.

The simplest $S_\alpha\pqty{s}$ we consider is inspired by the single qubit model with both dephasing and relaxation noise. In this case, $\Bqty{O_\alpha} \equiv \Bqty{X, Z} $ in the interaction Hamiltonian \eqref{eq:interaction_term}. In the adiabatic frame we have $\Bqty{S_\alpha\pqty{s}} \equiv \Bqty{Z\pqty{s}, X\pqty{s}}$, where
\bes
\label{eq:constructed_coupling}
\begin{align}
\label{eq:constructed_coupling_1}
    Z\pqty{s} &= \cos[\theta\pqty{s}]Z-\sin[\theta\pqty{s}]X \\
\label{eq:constructed_coupling_2}
    X\pqty{s} & = \cos[\theta\pqty{s}]X + \sin[\theta\pqty{s}]Z \ .
\end{align}
\ees
Furthermore, we assume the $\Bqty{\mathcal{B}_\alpha(s)}_\alpha$ are independent and identical with an Ohmic spectral density~\cite{albash_quantum_2012}: 
\begin{equation}
\label{eq:Ohmic}
    \gamma\pqty{\omega} = \frac{\eta g^2\omega e^{-\frac{\omega}{\omega_c}}}{1-e^{-\beta\omega}} \ ,
\end{equation}
where $\eta g^2$ is a dimensionless system-bath coupling constant and $\omega_c$ is the high-frequency cutoff.

\begin{figure*}[ht]
    \centering
        \subfigure{\includegraphics[width=\columnwidth]{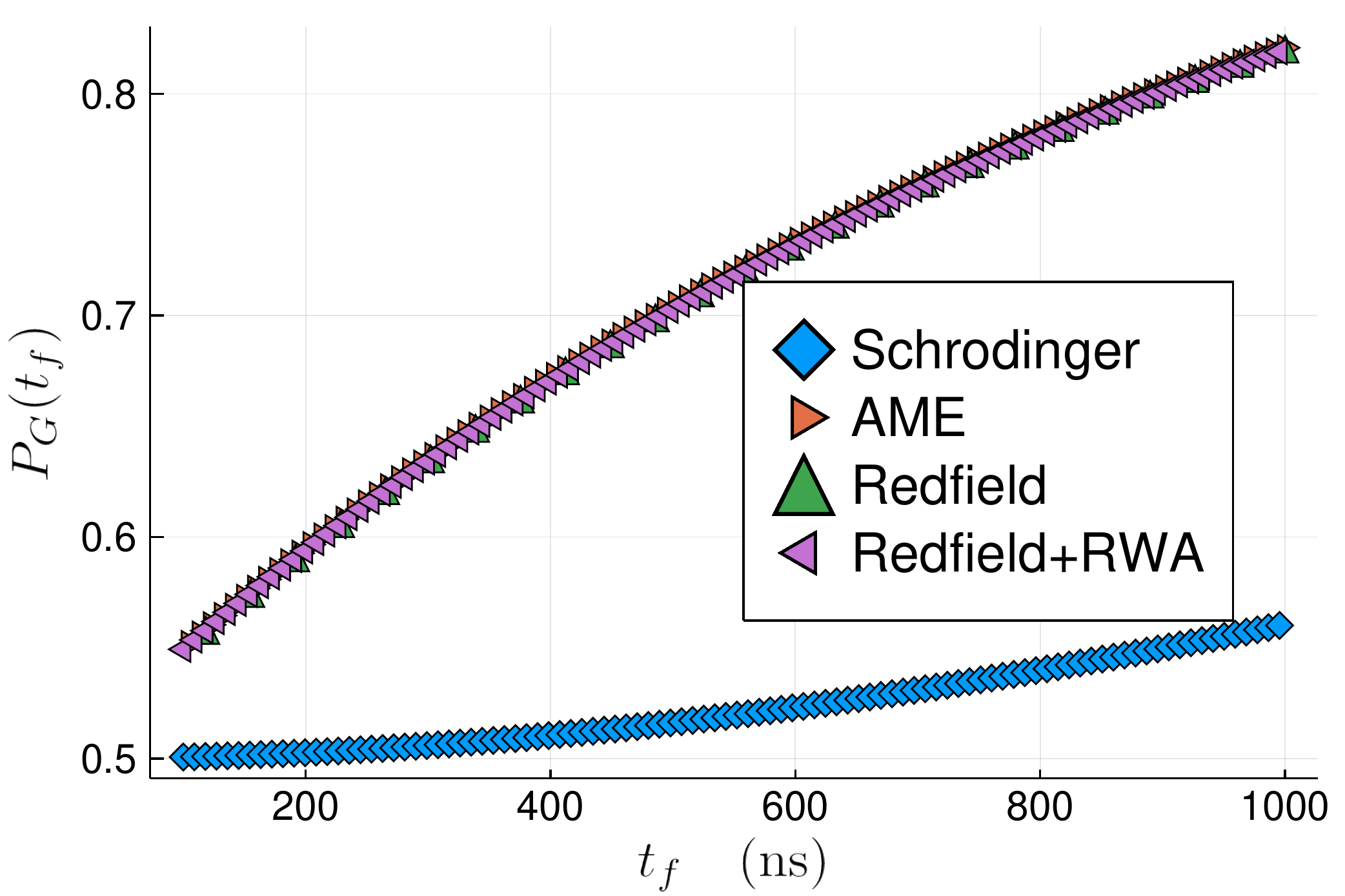}
        \label{fig:no_pause_evolution-a}}
    ~ 
\subfigure{\includegraphics[width=\columnwidth]{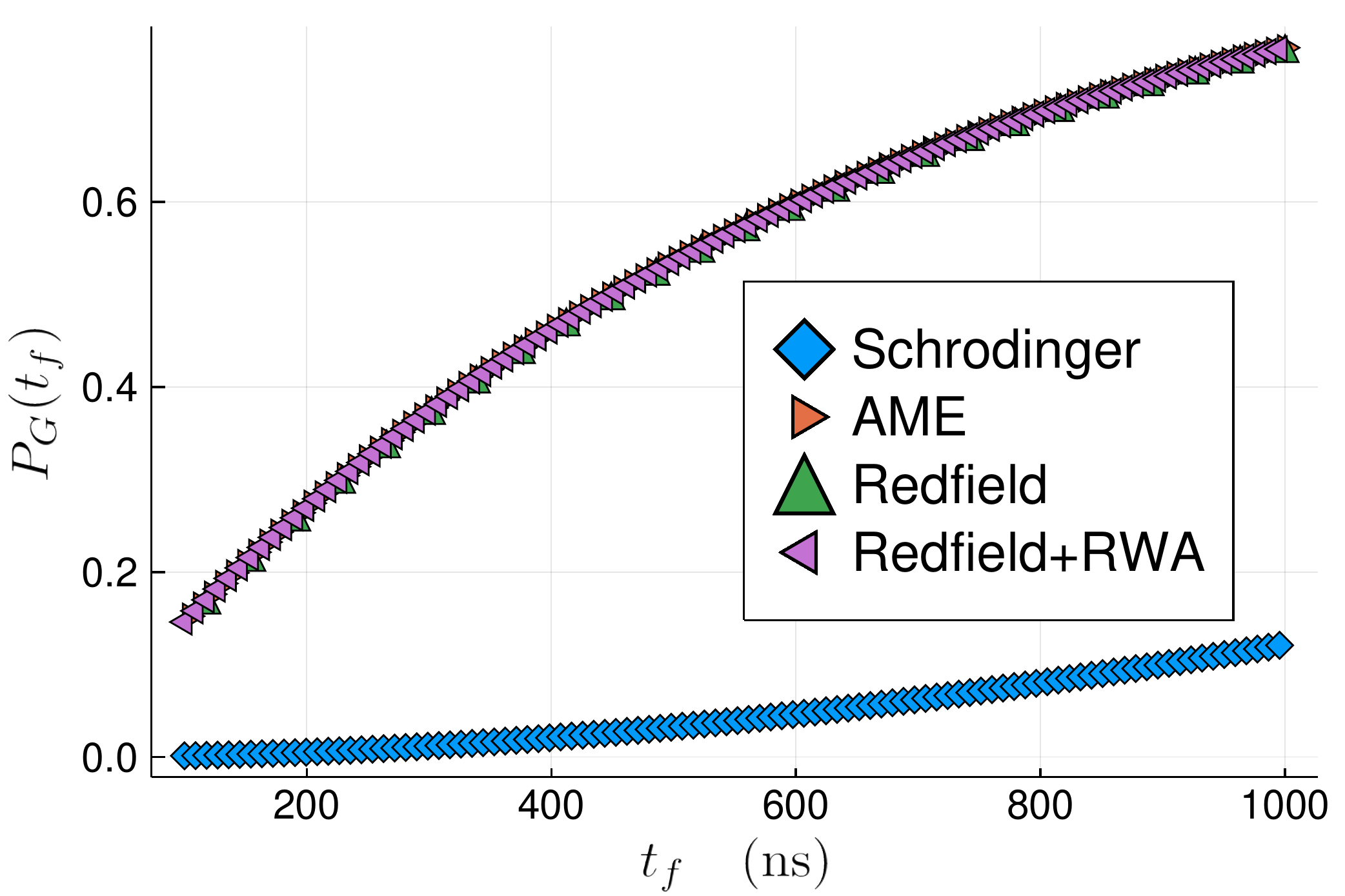}
        \label{fig:no_pause_evolution-b}}
    \caption{Success probability (without pausing) calculated in both closed and open system settings with boundary conditions: (a) $\theta\pqty{1} = \pi/2$; (b) $\theta\pqty{1}=\pi$. Open system simulations were done with all three variants of MEs described in Sec.~\ref{sec:open_system}, with an Ohmic bath spectral density [Eq.~\eqref{eq:Ohmic}]. Here and below the Ohmic bath parameters were chosen as typical of flux qubits (e.g., Refs.~\cite{albash_quantum_2012,dickson_thermally_2013,yan_flux_2016,Quintana:2017aa,Novikov:2018aa,khezri2020annealpath}): $2\pi \eta g^2 = 10^{-4}$, $T = 16$(mK) and $\omega_c/2\pi = 4$(GHZ). The results of the open system simulations overlap. Note the different vertical axis scales in (a) and (b).} 
    \label{fig:no_pause_evolution}
\end{figure*}

In Fig.~\ref{fig:no_pause_evolution}, results of the three variants of MEs described in Sec.~\ref{sec:open_system} are compared against the closed system case. Clearly, the relaxation present in the open system case drastically increases the ground state probability.
\begin{figure*}[ht]
    \centering
    \subfigure[]{\includegraphics[width=\columnwidth]{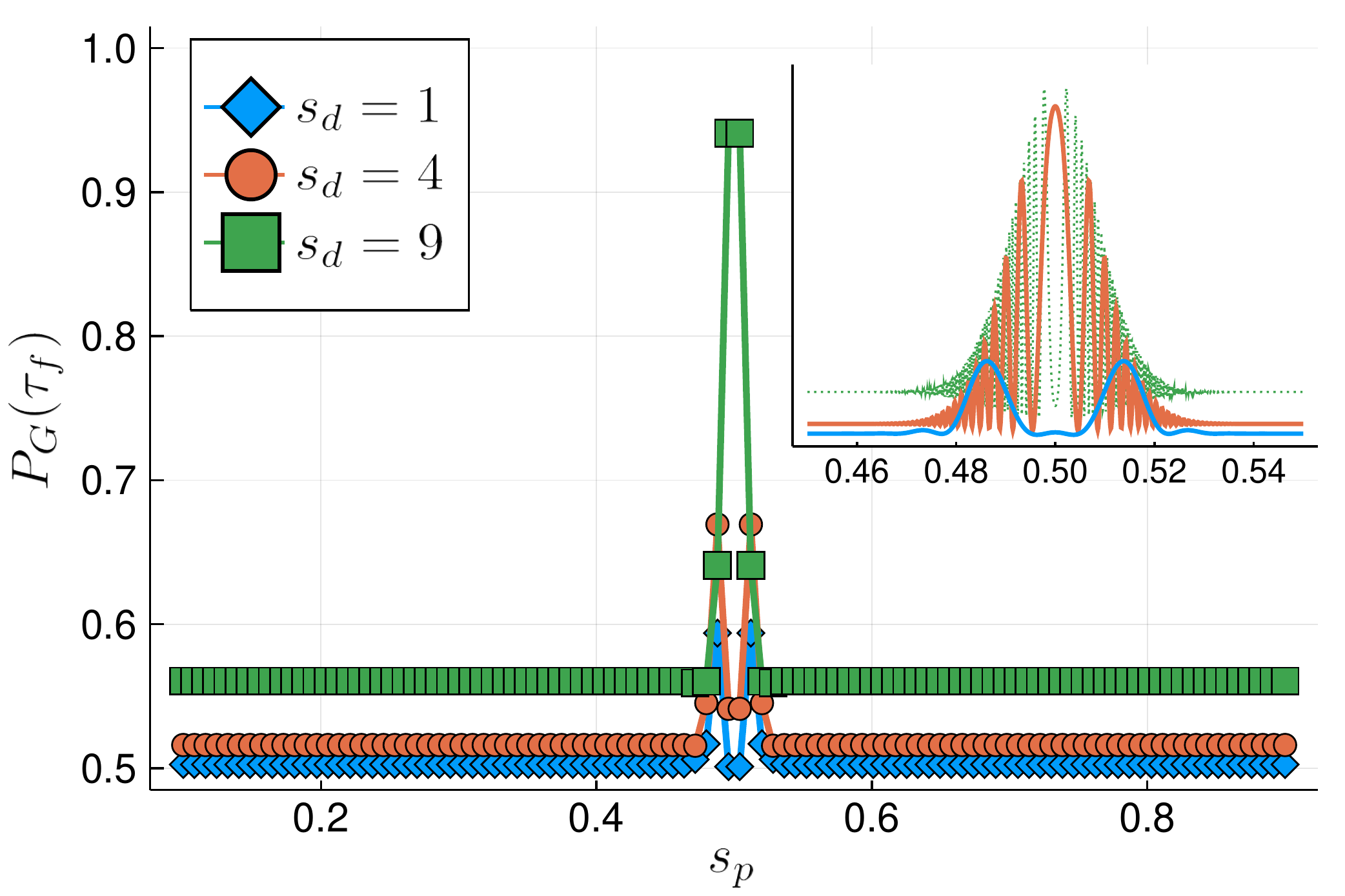}
        \label{fig:pausing_closed_system}}
    \subfigure[]{\includegraphics[width=\columnwidth]{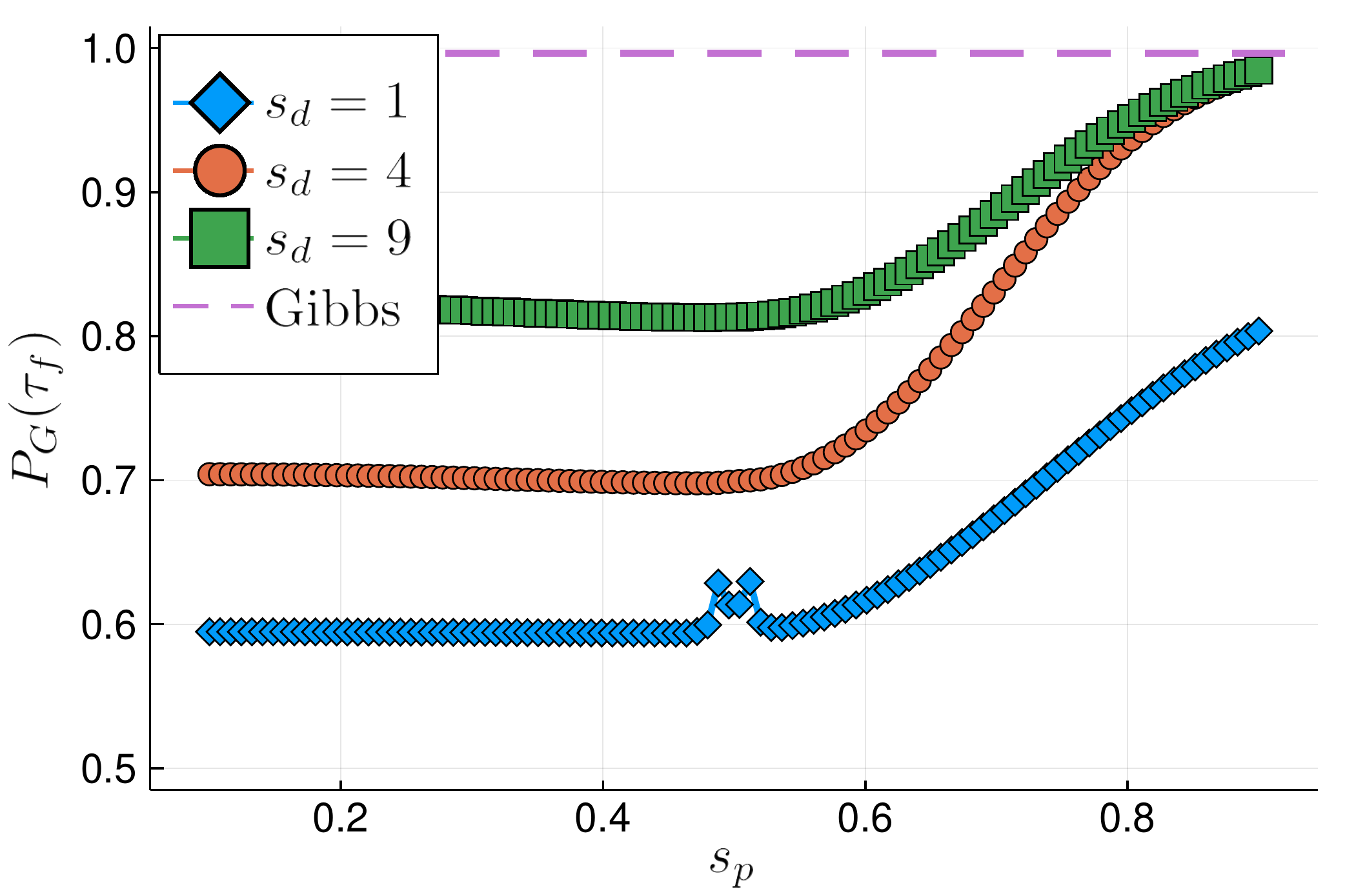}
    \label{fig:pausing_open_system}}\\
    
    \subfigure[]{\includegraphics[width=\columnwidth]{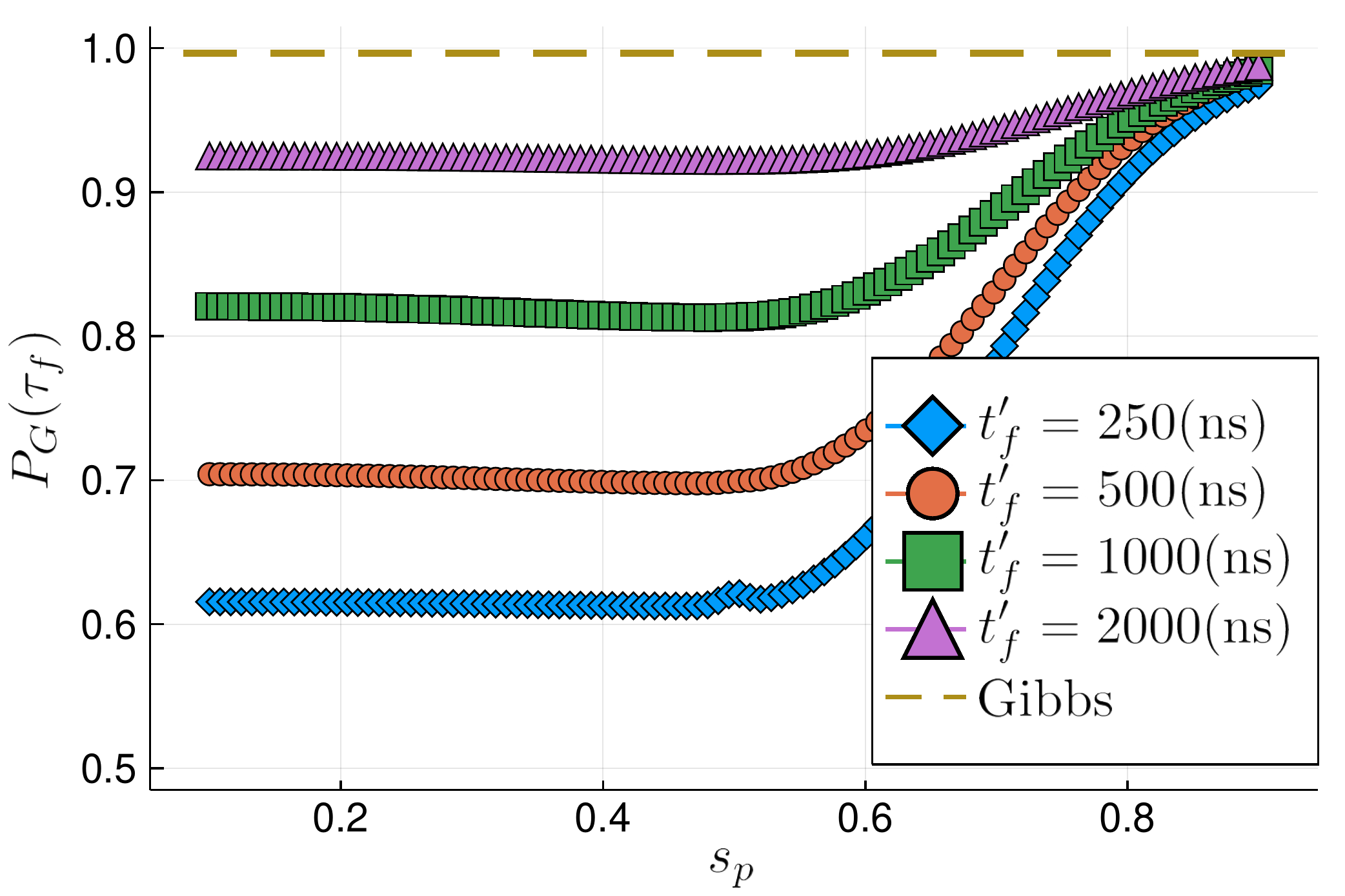}
    \label{fig:pausing_tf}}
    \subfigure[]{\includegraphics[width=\columnwidth]{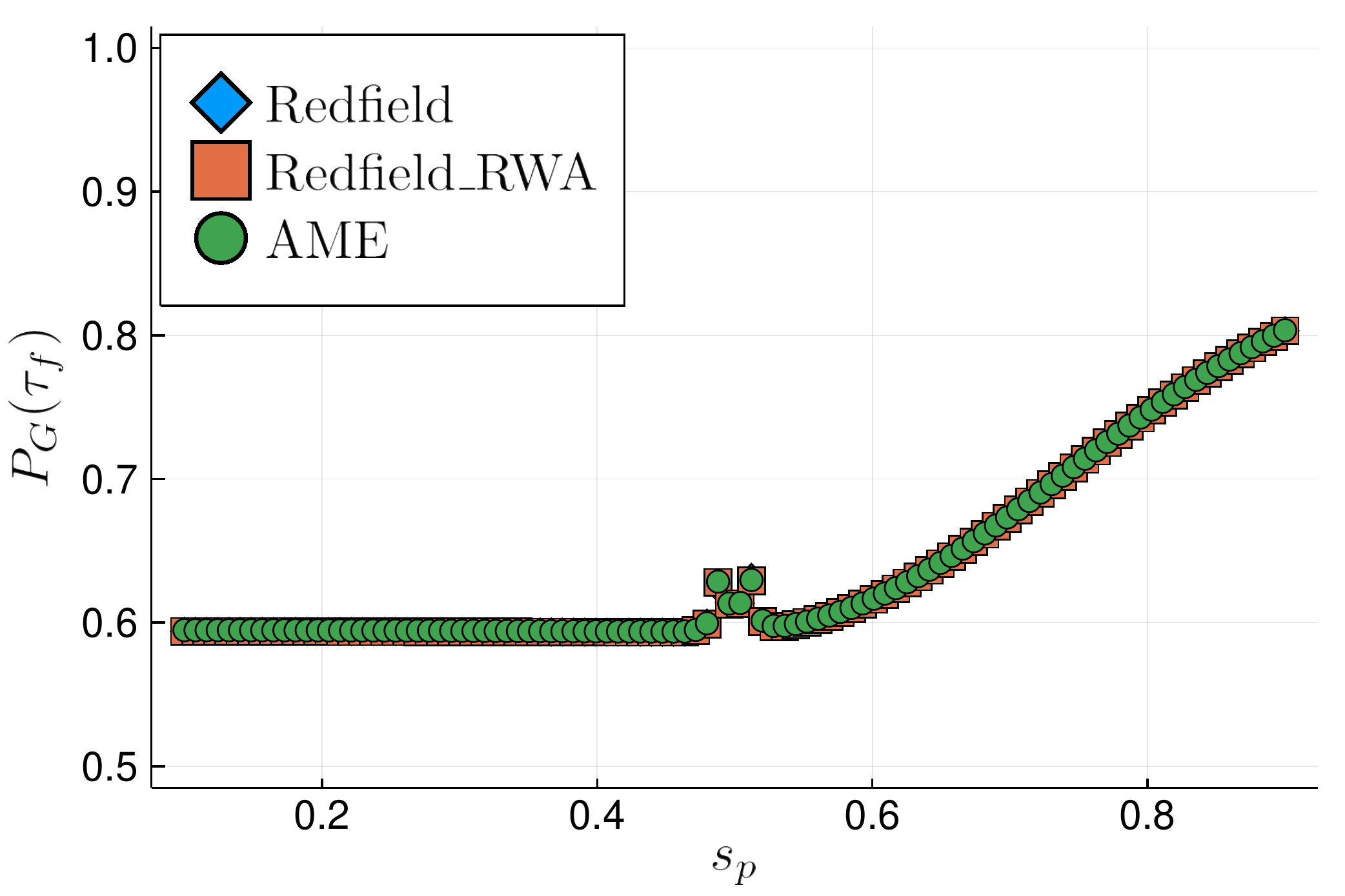}
    \label{fig:pausing_MEs}}
    \caption{Success probability for pausing schedules as shown in Fig.~\ref{fig:pausing}. The boundary condition for every plot is $\theta(0)=0$ and $\theta(1)=\pi/2$ (for $\theta(1)=\pi$, the plots are of similar shapes, with the lowest success probability being 0). For each curve the pausing duration $s_d$ is fixed and the pausing position $s_p$ is varied. (a) Closed system. The inset zooms in around $s_p=0.5$. (b) AME. (c) AME success probabilities for different $t'_f$ values with $s_d=4$. (d) Final success probabilities from three MEs with $s_d = 1$. For (a), (b) and (d), the original total annealing time $t_f$ is set to 100(ns) and the Ohmic bath parameters are chosen as: $2\pi \eta g^2 = 10^{-4}$, $T = 16$(mK) and $\omega_c/2\pi = 4$(GHZ). The dashed lines in (b) and (c) indicate the thermal ground state population of the final Hamiltonian.}
    \label{fig:adiabatic_pause}
\end{figure*}
In the parameter region we consider, all three variants of MEs give the same predictions. In principle, the adiabatic and RWA version of Redfield equation are unreliable in small gap problems~\cite{mozgunov_completely_2019}. However, in our construction the small gap/diabatic region is so narrow that any error introduced during that period can be safely ignored. This numerically justifies the delta function approximation we make in Sec.~\ref{sec:theoretical} below.

In addition, numerical results for a pausing schedule of the type shown in Fig.~\ref{fig:pausing} are presented in Fig.~\ref{fig:adiabatic_pause}. In these simulation we fixed the pausing duration $s_d$ and investigated the final success probability for different pausing positions $s_p$. Four primary observations arise from these results: (1) Increasing the total annealing time $t'_f$ improves the final success probability. This is because we are neither close to the adiabatic limit (closed system) nor to thermal equilibrium (open system). (2) There is a peak (or more precisely, oscillations) in the success probability if we pause right around the minimum gap. However, for the parameters chosen here this is mainly a closed system effect which is suppressed in the open system setting (we show Sec.~\ref{sec:search} below that the choice of the cutoff frequency matters a great deal). The oscillations can be understood as the interference pattern between two different paths leading to the final ground state~\cite{munoz-bauza_double-slit_2019}, because pausing splits the single region of diabatic evolution into two and effectively creates a Mach-Zehnder interferometer~\cite{Oliver:2005aa}. (3) Pausing only helps if it happens after the minimum gap. This phenomenon can be explained by our analytic model presented in Sec.~\ref{sec:theoretical}. (4) All the MEs still produce the same results in the presence of pausing, which suggests we may use the AME, since it has the simplest structure for pursuing analytical results.

\subsection{Theoretical Analysis}
\label{sec:theoretical}

We now present a theoretical analysis in support of the numerical results above, and to address the experimental findings in Ref.~\cite{marshall_power_2019}. Without loss of generality, we will assume $\theta(1) = \pi/2$ in the derivation. The proof also goes through for other boundary conditions. The first assumption we make is the localization of the geometric phase. To be more specific, we assume
\begin{equation}
    \dot{\theta}\pqty{s} \approx 0 \quad \forall s \notin \bqty{\mu_\theta-c\alpha_\theta, \mu_\theta+ c\alpha_\theta}  
\end{equation}
where $c$ is a dimensionless constant and $c\alpha_\theta$ measure the effective width of the Gaussian pulse. Later, we will take the limit $\alpha_\theta \to 0$. This condition holds for all the hard instances we consider in this work. In the small $\alpha_\theta$ limit, the Redfield Eq.~\eqref{eq:adiabatic_tcl2} can be treated separately inside/outside the region $\bqty{\mu_\theta-c\alpha_\theta, \mu_\theta+ c\alpha_\theta}$. Within this Landau-Zener (LZ) region, the geometric phase dominates and all the other terms can be considered as a perturbation. With the detailed derivation given in Appendix~\ref{app:beam-splitter}, we prove that, in the limit of weak coupling and small $\alpha_\theta$, the evolution across the LZ region can be approximated by a ``diabatic pulse" unitary of the following form
\begin{equation}
\label{eq:U}
    U=
    \begin{pmatrix}
    \cos(\varphi) &
    -i \sin(\varphi)\\
    -i \sin(\varphi)&
    \cos(\varphi) 
    \end{pmatrix} \ ,
\end{equation}
where
\begin{equation}
\label{eq:varphi}
\varphi = \frac{\pi}{4} e^{-(t_f/t_{\mathrm{ad}})^2} \ , \quad    t_{\mathrm{ad}} = \frac{\sqrt{2}}{\alpha_\theta \int_0^{\mu_\theta} \Omega\pqty{s} \dd{s}} \ .
\end{equation}
In the limit $\alpha_\theta \to 0$ the closed-system adiabatic time scale $t_{\mathrm{ad}}$ diverges. This is an approximation to real computational (small gap) problems where the adiabatic time scale is infinite for practical purposes. 

Outside the LZ region, $\dv{\theta}{\tau}\approx 0$ and hence the Hamiltonian \eqref{eq:pause_H} can be written as
\begin{equation}
    H(\tau) = -\frac{t'_f}{2}\Omega(\tau) Z + t'_f\sum_\alpha g_\alpha S_\alpha\pqty{\tau}\otimes\mathcal{B}_\alpha + H_{\mathrm{B}}\ ,
\end{equation}
We also assume the position of the pause is outside the LZ region:
\begin{equation}
s_p \notin \bqty{\mu_\theta^- , \mu_\theta^+} , \qquad \mu_\theta^\pm\equiv\mu_\theta\pm c\alpha_\theta \ .
\end{equation}
This is in accordance with the experimental protocol of Ref.~\cite{marshall_power_2019}, where pausing was found to be effective past the position of the avoided crossing, and is explained theoretically below in terms of the absence of a pausing effect before $s=\mu_\theta$.

Following Ref.~\cite{albash_decoherence_2015}, the AME can be written as two fully decoupled parts, which simplifies the derivation compared to the other master equations considered above: 
\bes
\begin{align}
    \dot{\rho}_{00}(\tau) &= -\dot{\rho}_{11}(\tau) =t'_f \Gamma_{01}(\tau)\rho_{11}(\tau) - t'_f\Gamma_{10}(\tau)\rho_{00}(\tau) \label{eq:diag}\\
    \dot{\rho}_{01}(\tau) & = -it'_f\pqty{\omega_{01}(\tau)+\Sigma_{01}(\tau)}\rho_{01}(\tau) - \xi_{01}(\tau) \rho_{01}(\tau)
\end{align}
\ees
where
\bes
\begin{align}
    \omega_{01}(\tau) &= -\omega_{10}(\tau) = -\Omega(\tau)\\
    \label{eq:Gamma01tau}
    \Gamma_{01}(\tau) &= \sum_\alpha \gamma_\alpha\pqty{\Omega(\tau)}\abs{S_\alpha^{01}(\tau)}^2 = e^{\beta \Omega(\tau)}\Gamma_{10}(\tau) \\
    \Sigma_{01}(\tau) &= -\Sigma_{10}(\tau)= \sum_\alpha \abs{S_\alpha^{01}}^2(\mathcal{S}(\omega_{01})-\mathcal{S}(\omega_{10})) \\
    \xi_{01}\pqty{\tau} &= \frac{1}{2}\pqty{\Gamma_{01}+\Gamma_{10}} + \frac{1}{2}\sum_\alpha \gamma_\alpha\pqty{0}\abs{S^{00}_\alpha + S^{11}_\alpha} 
\end{align}
\ees
Our primary interest is in the ground state population [Eq.~\eqref{eq:diag}]. It can be rewritten as
\bes
\begin{align}
\label{eq:ground_pop_eq}
    \dot{\rho}_{00}(\tau) &= \Gamma(\tau)\bigg[1-\pqty{1+e^{-\beta\Omega(\tau)}} \rho_{00}(\tau)\bigg]\\
\label{eq:dimless-rate}
    \Gamma(\tau) &\equiv t'_f \Gamma_{01}(\tau)\ ,
\end{align}
\ees
where $\Gamma(\tau)$ is the dimensionless relaxation rate subject to pausing.\footnote{Recall our convention of denoting by $Q(s)$ the original quantity and by $Q(\tau)$ the corresponding paused quantity. In the case of $\Gamma(s)$ this includes the paused anneal time $t'_f$, so in that sense it represents a mixed quantity.}
The off-diagonal elements of the density matrix decay exponentially with a rate determined by $\xi_{01}(\tau)$. As a consequence, if we start in the ground/thermal state of the initial Hamiltonian, right before the LZ region, the system density matrix will to an exponentially good approximation have only diagonal elements:
\begin{equation}\label{eq:before_pulse}
    \rho(\mu_\theta^-) = \mqty(P(\mu^-_\theta) & 0 \\ 0 & 1-P(\mu^-_\theta)) \ .
\end{equation}
After crossing the LZ region, using Eq.~\eqref{eq:U} the state becomes
\bes
\label{eq:after_pulse}
\begin{align}
    \rho(\mu_\theta^+) &= \mqty(P_\varphi& -i(P_0-P_1)\sin2\varphi \\ i(P_0-P_1)\sin2\varphi & 1-P_\varphi)\\
    P_0 &= P(\mu_\theta^-) = 1- P_1 \\
    P_\varphi  &= P_0 \cos^2{\varphi} + P_1\sin^2{\varphi} \ .
    \label{eq:Pphi}
\end{align}
\ees
Note that $P_\varphi$ is the ground state population right after the minimum gap is crossed.
We can see from Eqs.~\eqref{eq:before_pulse} and~\eqref{eq:after_pulse} that pausing before $\mu_\theta$ has no effect on the final results. Therefore, in the following discussion, we will assume the pausing position is after the diabatic pulse: $s_p>\mu_\theta$. The solution of Eq.~\eqref{eq:ground_pop_eq} from $\mu^+_\theta$ to $\tau_f= 1 + s_d$
can be written as
\bes
\label{eq:population}
\begin{align}
    \rho_{00}(\tau_f) &= \exp[-\int_{\mu^+_\theta}^{\tau_f}\dd{\tau}\pqty{1+e^{-\beta \Omega(\tau)}}\Gamma(\tau)] \nonumber\\
    &\times\Bigg\{P_\varphi+\int_{\mu^+_\theta}^{\tau_f}\dd{\tau}\Gamma(\tau) 
    \\    &\times
    \exp[\int_{\mu^+_\theta}^{\tau}\dd{\tau'}\pqty{1+e^{-\beta\Omega(\tau')}}\Gamma(\tau')]\Bigg\} \nonumber\\
    & = \mathcal{F}_d(\tau_f, s_p, s_d) + \mathcal{F}_g(\tau_f, s_p, s_d)\ ,
\end{align}
\ees
where
\bes
\begin{align}
\mathcal{F}_d(\tau_f, s_p, s_d) &= P_\varphi G(\mu^+_\theta)  
\label{eq:F_d}\\
    \mathcal{F}_g(\tau_f, s_p, s_d) &= \int_{\mu^+_\theta}^{\tau_f}\dd{\tau}\Gamma(\tau) G(\tau) \ ,
\label{eq:F_g}
\end{align}
\ees
and where
\bes
\begin{align}
\label{eq:G}
    G(\tau) &= \exp\bigg(-\int_{\tau}^{\tau_f}\dd{\tau'}X(\tau')\bigg) \\
\label{eq:X} 
    X(\tau) &= \bigg(1+e^{-\beta\Omega(\tau)}\bigg)\Gamma(\tau) \ .
\end{align}
\ees

Note that the functions $\Omega(\tau)$ and $\Gamma(\tau)$ have an implicit dependence on $s_p$ and $s_d$:
\bes
\begin{align}
    \label{eq:Omegatau}
    \Omega(\tau) &= \Omega(s(\tau)) \\
    \label{eq:Gammatau}
    \Gamma(\tau) &= (1+s_d)t_f\sum_\alpha \gamma_\alpha[\Omega(s(\tau))]\abs{S_\alpha^{01}(s(\tau))}^2 ,
\end{align}
\ees
where we combined Eqs.~\eqref{eq:Gamma01tau} and ~\eqref{eq:dimless-rate}.
To achieve the maximum success probability, we need to solve the following optimization problem:
\begin{equation}
\label{eq:optimization}
    \argmax_{\Bqty{s_p, s_d}} \mathcal{F}_d(\tau_f, s_p, s_d) + \mathcal{F}_g(\tau_f, s_p, s_d) \ .
\end{equation}

\subsection{Numerical evidence for an optimal pausing position}
\label{sec:search}

\begin{figure*}[t]
    \centering
         \subfigure{\includegraphics[width=.32\textwidth]{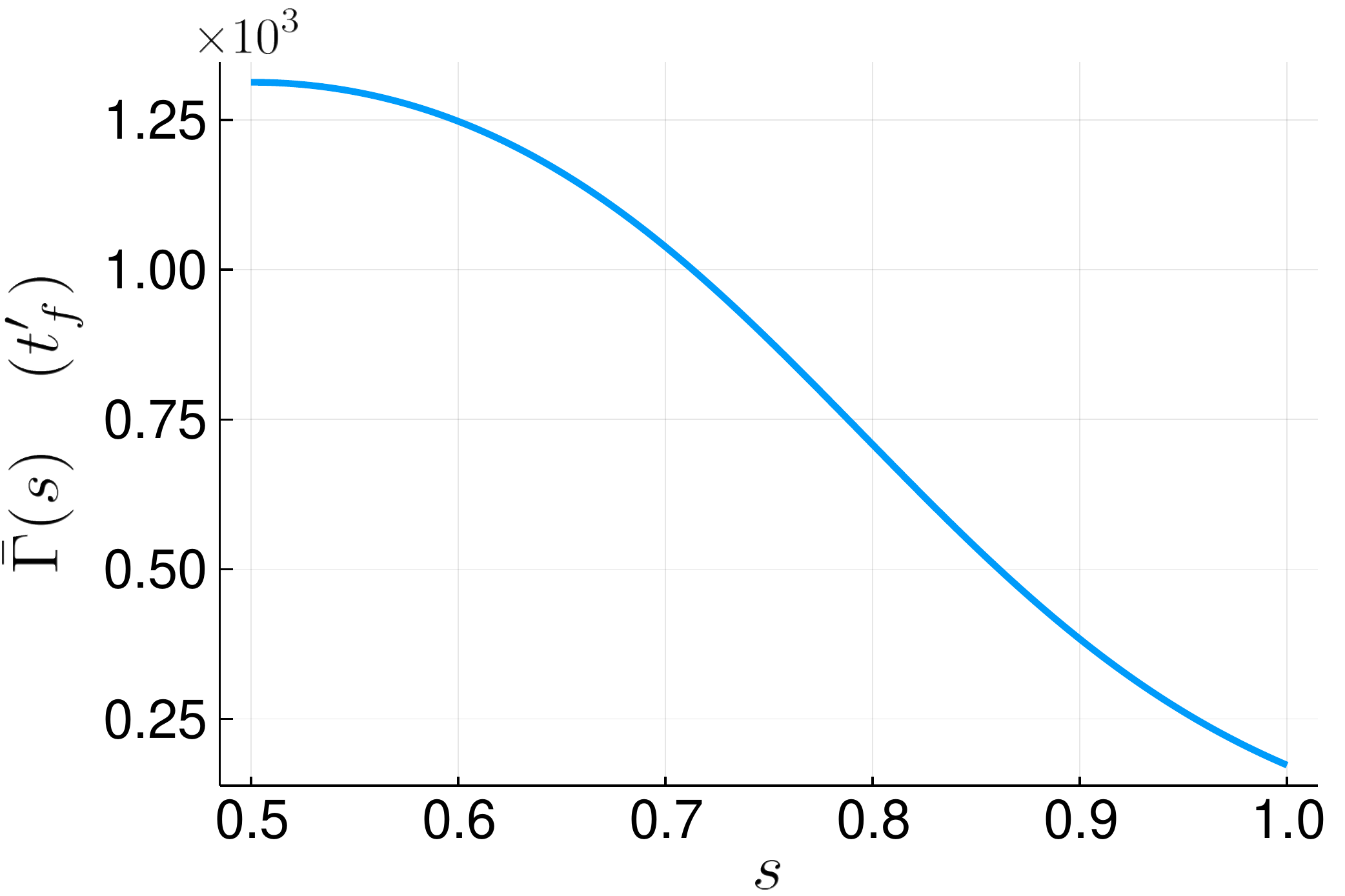}
        \label{fig:monotonic_decreasing}}
    ~ 
        \subfigure{\includegraphics[width=.32\textwidth]{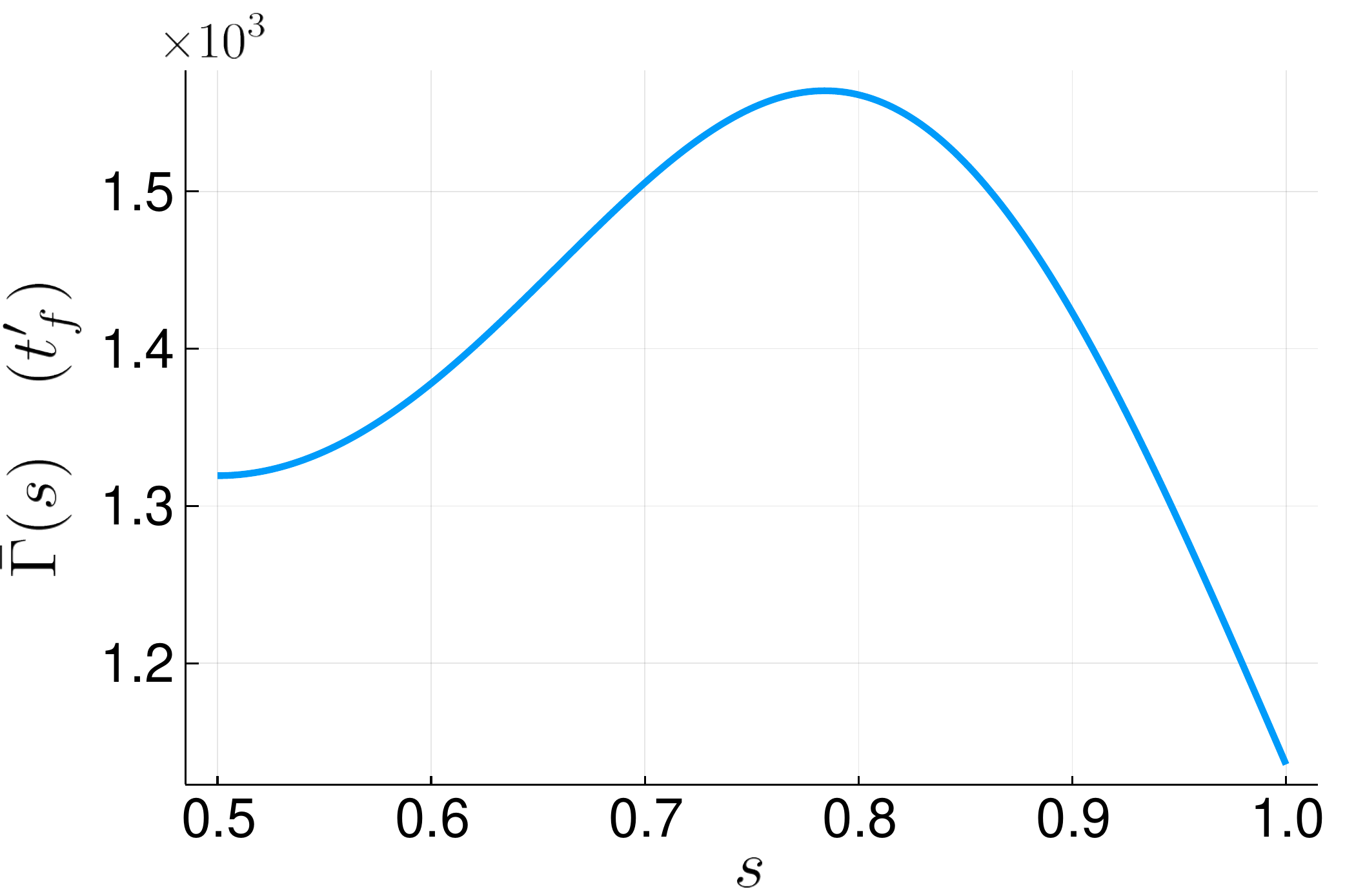}
        \label{fig:U_shape}}
        \subfigure{\includegraphics[width=.32\textwidth]{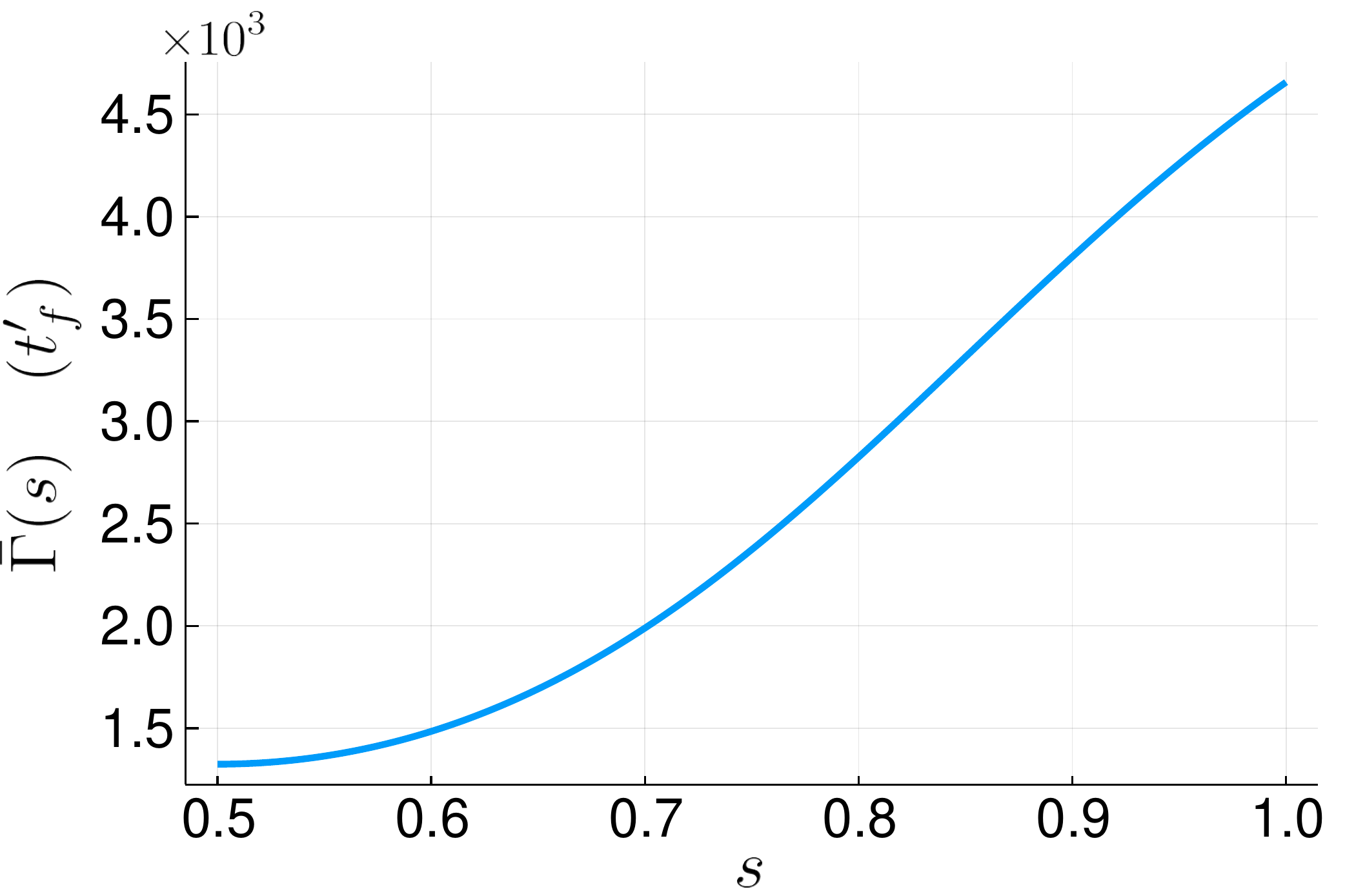}
        \label{fig:monotonic_increasing}}
    \caption{$\bar{\Gamma}(s)$ [the dimensionless relaxation rate in Eq.~\eqref{eq:Gamma01tau} as a function of $s$, i.e., with $\tau$ replaced by $s$] for an Ohmic bath with different cutoff frequencies: $\omega_c/2\pi = 0.5$ (GHz), monotonically decreasing $\bar{\Gamma}(s)$ (left); $\omega_c/2\pi = 1$ (GHz), non-monotonic (middle); $\omega_c/2\pi = 4$ (GHz), monotonically increasing $\bar{\Gamma}(s)$ (right).}
    \label{fig:Gamma_s}.
\end{figure*}
\begin{figure*}[t]
    \centering
        \subfigure{\includegraphics[width=.32\textwidth]{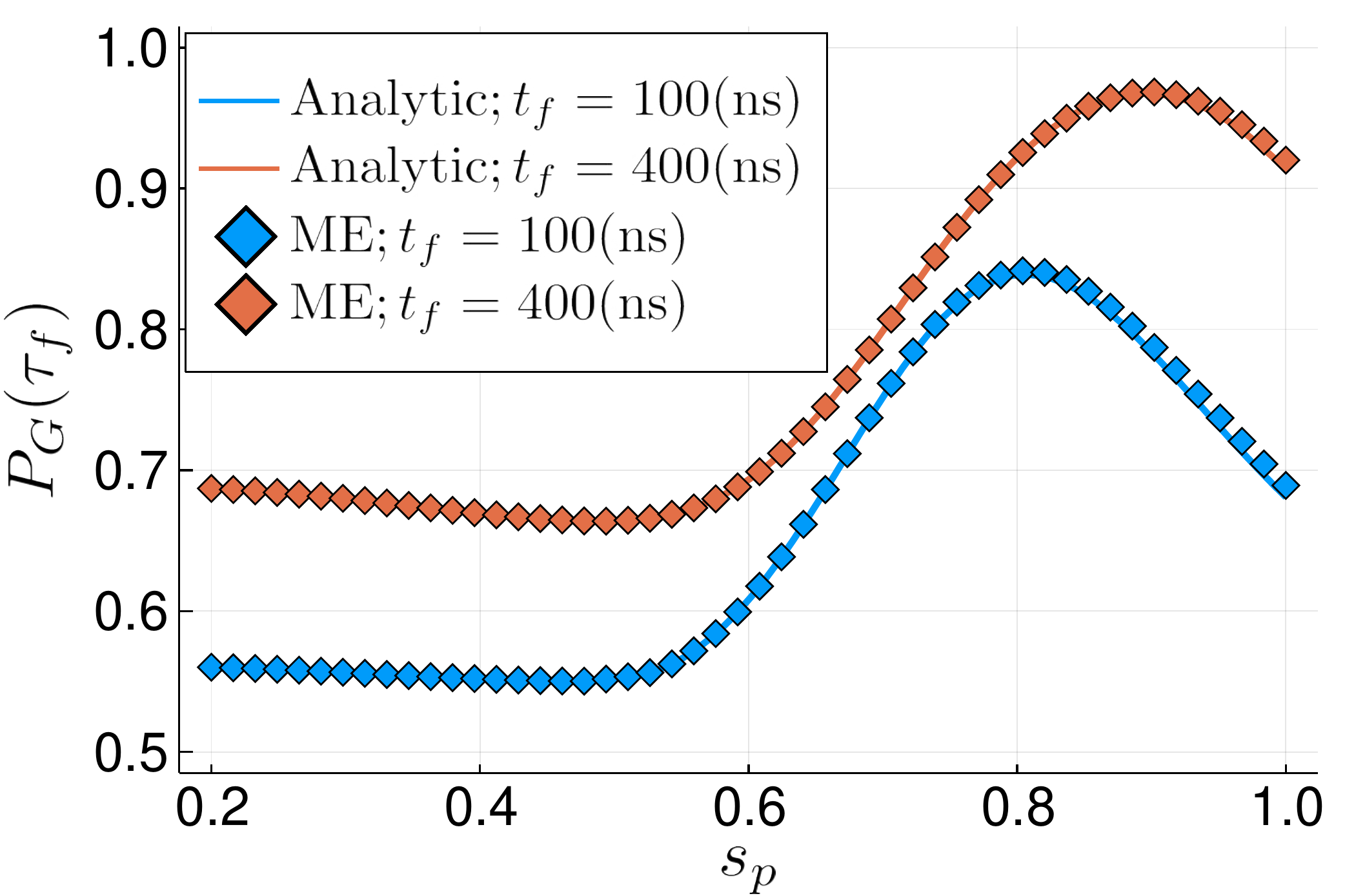}
        \label{fig:optimal_position_decreasing}}
        \subfigure{\includegraphics[width=.32\textwidth]{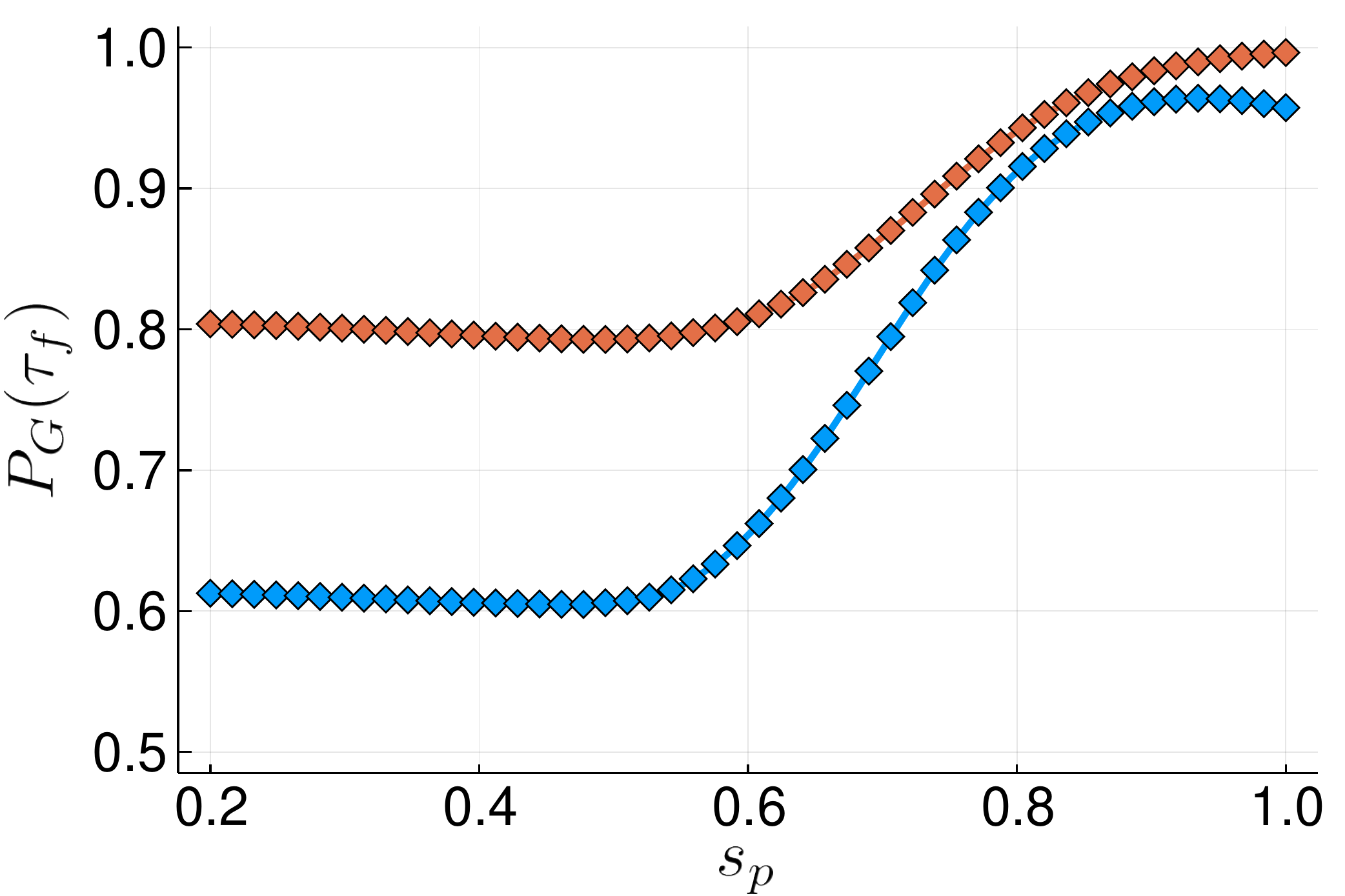}
        \label{fig:optimal_position_U}}
        \subfigure{\includegraphics[width=.32\textwidth]{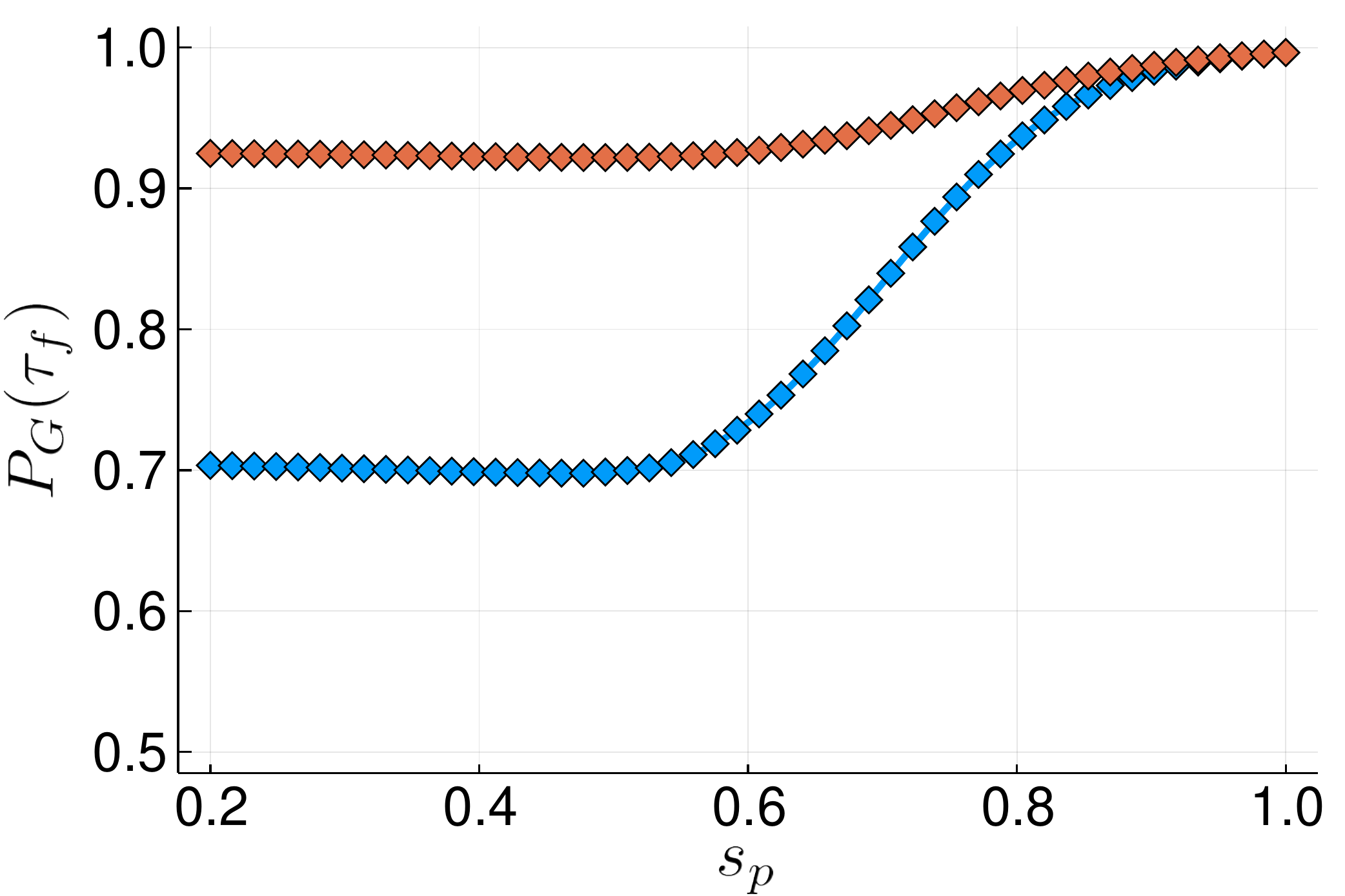}
        \label{fig:optimal_position_increasing}}
    \caption{Success probability \textit{vs} pausing position for the three Ohmic bath cases shown in Fig.~\ref{fig:Gamma_s}, in the same order from left to right. The existence of a maximum in the left and middle panels shows that an optimal pausing exists for the corresponding decay rates. The qubit frequency $\Omega(s)$ [Eq.~\eqref{eq:single_qubit_adiabatic_hamiltonian}] changes between $1.88 \mathrm{GHz}$ and $4.77 \mathrm{MHz}$ during the anneal. 
Other parameters are: $\eta g_z^2 = \eta g_y^2 = 10^{-4} /2 \pi$ and $T = 16$(mK). The analytic solution refers to Eq.~\eqref{eq:population}. For master equation simulations, the schedules are chosen according to Fig.~\ref{fig:schedules}. The pausing duration is fixed at $s_d = 4$. All three panels share the same legend.}
    \label{fig:optimal_position}
\end{figure*}

First, noticing that the quantities defined in Eqs.~\eqref{eq:X},~\eqref{eq:Omegatau}, and~\eqref{eq:Gammatau} can be functions of either $s$ (unpaused) or $\tau$ (paused), to avoid any ambiguity \emph{we henceforth use the notation $\bar{Q}(s)$ when the argument is $s$.} For example, $\bar{\Gamma}(1)$ means $\Gamma(s=1)$ instead of $\Gamma(\tau=1)$. Note that this modifies our previous convention of denoting by $Q(s)$ the original quantity and by $Q(\tau)$ the corresponding paused quantity.

Then, we analyze the optimal pausing position $s_p^*$, for a given pausing duration $s_d$. We  consider an Ohmic bath with different cutoff frequencies. This leads to different behaviors of the dimensionless relaxation rate $\bar{\Gamma}(s)$, from decreasing to non-monotonic to increasing, as illustrated in Fig.~\ref{fig:Gamma_s}. Before proceeding, we emphasize that, as is clear from Eq.~\eqref{eq:Gammatau}, the monotonicity properties of $\bar{\Gamma}(s)$ depend on three different factors:
\begin{itemize}
    \item The gap of the projected TLS
    \item The projected system-bath coupling operators [Eq.~\eqref{eq:proj-SB-ops}]
    \item The noise spectrum
\end{itemize}
In our examples, the first two are fixed by Eqs.~\eqref{eq:gaussian_gap} and~\eqref{eq:constructed_coupling} respectively. Thus, at a given temperature, what remains is only the noise spectrum, or more specifically the only tuning parameter for the monotonicity  of $\bar{\Gamma}(s)$ is the Ohmic bath cutoff frequency $\omega_c$. The relationship between $\omega_c$ and the monotonicity of $\gamma(\omega)$ within the qubit frequency range, which is sufficient to determine the monotonicity of $\bar{\Gamma}(s)$, is not straightforward. Thus the values in Fig.~\ref{fig:Gamma_s} were chosen after numerical investigation. More generally, the first two factors above are of equal importance to the noise spectrum. For example, in the $16$-qubit problem of Ref.~\cite{dickson_thermally_2013} with pure dephasing couplings, the strength of the projected system-bath coupling operators $\sum_\alpha\abs{S_\alpha^{01}(s(\tau))}^2$ plays a key role as well: it has a peak around the minimum gap region and decreases to almost zero afterwards.

The corresponding success probabilities as a function of pausing position obtained by numerically solving the AME and via the analytic expression~\eqref{eq:population} are shown in Fig.~\ref{fig:optimal_position}, and are in excellent agreement. There are two main observations: 
\begin{itemize}
\item An optimal pausing position exists in the middle of the anneal for the cases illustrated in Fig.~\ref{fig:monotonic_decreasing} and Fig.~\ref{fig:U_shape} where $\bar{\Gamma}(s)$ is not monotonically increasing.
\item When $\bar{\Gamma}(s)$ is monotonically increasing after the minimum gap, it is always better to (trivially) pause near the end of anneal.
\end{itemize}

Guided by these numerical results, we next prove that a non-trivial optimal pausing position exists provided that the dimensionless relaxation rate given in Eq.~\eqref{eq:dimless-rate} (i.e., comparing the actual relaxation rate $\Gamma_{01}$ to the anneal time $t_f'$) is monotonically decreasing, with respect to $s$, from the end of the avoided crossing to the end of the anneal.

\subsection{Existence of optimal pausing point}
\label{sec:existence}

In this section we prove the following theorem, which provides sufficient conditions for the existence of a non-trivial optimal pausing position, thus generalizing and formalizing the numerical evidence exhibited above. 

\begin{mytheorem}[Optimal pausing point]
\label{thm:opt-pausing}
Let $\bar{\Gamma}(s) = (1+s_d)t_f \bar{\Gamma}_{01}(s)$ be the dimensionless relaxation rate, let the instantaneous ground state probability after the minimum gap is crossed be denoted $P_\varphi$ [Eq.~\eqref{eq:Pphi}], and let $\bar{P}_{\text{th}}\pqty{s} = \frac{1}{1+e^{-\beta\bar{\Omega}\pqty{s}}}$ be the thermal ground state probability at $s$.

There exists a non-trivial optimal pausing point $s_p^* \in \pqty{\mu_\theta^+, 1}$ for a fixed pausing duration $s_d$, i.e., 
\begin{equation}\label{eq:app_derivative_0}
    \pdv{\rho_{00}}{s_p} \Big\vert_{s_p = s^*_p} = 0 \ ,
\end{equation}
if  
\begin{enumerate}
\item The dimensionless relaxation rate decreases at the boundaries of the interval $\bqty{\mu_\theta^+, 1}$, i.e., $
\bar{\Gamma}'\pqty{\mu_\theta^+}<0$ and $\bar{\Gamma}'\pqty{1}<0$.%
\footnote{Note that henceforth, for notational simplicity, we use the prime symbol for the derivative with respect to $s$ throughout this work.}
\item The dimensionless relaxation rate is large right after the minimum gap: $s_d \bar{\Gamma}\pqty{\mu_\theta^+} > 
c_1$ where $c_1=O(1)$.
\item The dimensionless relaxation rate is small at the end of the anneal: $s_d\bar{\Gamma}(1) < c_2$ where $c_2=O(1)$.
\item The ground state population at the end of the anneal is subthermal: \\
$1-(1-P_\varphi)e^{-\int_{\mu_\theta^+}^{1}\bar{\Gamma}\pqty{s}\dd{s}} \leq \bar{P}_{\text{th}}(1)$.
\end{enumerate}
\end{mytheorem}

We remark that the existence of a non-trivial optimal pausing point is possible under a substantially broader set of conditions than implied by Theorem~\ref{thm:opt-pausing}, as is shown in our proof below. The Theorem states a simplified set of conditions for ease of presentation and interpretation. The reader who is not interested in the technical details of the proof may skip ahead to the conclusions in Sec.~\ref{sec:conc}.

\subsection{Proof of Theorem~\ref{thm:opt-pausing}}

The proof follows.
Because $\pdv{\rho_{00}}{s_p}$ is a continuous function of $s_p$, a sufficient condition for Eq.~\eqref{eq:app_derivative_0} is
\begin{equation}
    \pdv{\rho_{00}}{s_p} \Big\vert_{s_p = \mu^+_\theta} > 0 \quad \mathrm{and} \quad \pdv{\rho_{00}}{s_p} \Big\vert_{s_p = 1} < 0\ .
    \label{eq:46}
\end{equation}
We thus divide the proof into two parts, one for each of these two inequalities.

\subsubsection{Proof of $\pdv{\rho_{00}}{s_p} \Big\vert_{s_p = \mu^+_\theta} = \pdv{\mathcal{F}_g}{s_p}\Big\vert_{s_p = \mu^+_\theta} + \pdv{\mathcal{F}_d}{s_p} \Big\vert_{s_p = \mu^+_\theta} > 0$}

Consider first the $\mathcal{F}_d$ term [Eq.~\eqref{eq:F_d}]:
\beq
\mathcal{F}_d = 
P_\varphi \exp[-\int_{\mu^+_\theta}^{\tau_f}\dd{\tau}X(\tau)]  \ .
\eeq 
The partial derivative at $s_p = \mu_\theta^+$ is
\begin{equation}
     \pdv{\mathcal{F}_d}{s_p} \Big\vert_{s_p=\mu^+_\theta} = -P_\varphi s_d\bar{X}'(\mu_\theta^+) G(\mu_\theta^+) \ ,
     \label{eq:58}
\end{equation}
where the factor of $s_d$ arises from $\int_{\mu^+_\theta}^{\tau_f}\pdv{X\pqty{\tau}}{s_p}\Big\vert_{s_p=\mu_\theta^+} \dd{\tau} = \bar{X}'(\mu_\theta^+) \int_{\mu^+_\theta}^{\mu_\theta^+ + s_d} 1 d\tau$. 
The relation between $\partial X / \partial s_p$ and $\bar{X}'$ is detailed in Appendix~\ref{app:dFg}. 

Likewise, the partial derivative at $s_p = 1$ is
\begin{equation}
\label{eq:app_dFd_1}
     \pdv{\mathcal{F}_d}{s_p} \Big\vert_{s_p=1} = -P_\varphi s_d\bar{X}'(1) G(\mu_\theta^+) \ .
\end{equation}

As for $\mathcal{F}_g$, again in Appendix~\ref{app:dFg} we derive the following identities:
\bes
\label{eq:app_dFg}
\begin{align}
\label{eq:app_dFg_x}
&    \pdv{\mathcal{F}_g}{s_p} \Big\vert_{s_p=\mu_\theta^+} =  s_d G\pqty{\mu_\theta^+}\int_{0}^{1}\dd{x} e^{\bar{S}x} \bigg(\bar{\Gamma}'\pqty{\mu_\theta^+}   \\
 &\quad     -s_d\bar{\Gamma}\pqty{\mu_\theta^+} \bar{X}'(\mu_\theta^+)(1-x)\bigg) \ , \quad \bar{S}\equiv s_d\bar{X}\pqty{\mu_\theta^+} \notag \\
    \label{eq:app_dFg_4}
&    \pdv{\mathcal{F}_g}{s_p} \bigg\vert_{s_p=1} = -s_d \bar{X}'\pqty{1}\int_{\mu^+_\theta}^{1}\dd{\tau}G\pqty{\tau}\Gamma\pqty{\tau} \\
    &\quad + \int_{1}^{1+s_d}\dd{\tau}G\pqty{\tau}\Big(\bar{\Gamma}'(1) -\bar{\Gamma}(1)\bar{X}'(1)\pqty{s_d+1-\tau} \Big) \notag\ 
\end{align}
\ees

To make further progress we now assume that:
\begin{myassumption}
\label{as:1}
$\bar{\Gamma}(s)$ is decreasing at the boundaries of the interval $\bqty{\mu_\theta^+, 1}$,\footnote{These two assumptions are deduced from our observations of examples we have studied numerically. In principle, one could choose a different set of assumptions and perform a similar  analysis. The final result will be different from Theorem~\ref{thm:opt-pausing}.} i.e.,
\beq
\bar{\Gamma}'\pqty{\mu_\theta^+}<0\ , \quad \bar{\Gamma}'\pqty{1}<0\ .
\eeq 
\end{myassumption}

Note that $\bar{\Omega}'(\mu_\theta^+)>0$ since the gap grows for $s> \mu_\theta$ [recall Eq.~\eqref{eq:gaussian_gap} and that we assumed $\mu_\theta \approx \mu_g$]. As a consequence, upon taking the derivative of Eq.~\eqref{eq:X} we obtain 
\beq
{\bar{X}}' = -\beta \bar{\Omega}' e^{-\beta \bar{\Omega}}\bar{\Gamma} + {\bar{\Gamma}}'(1+e^{-\beta \bar{\Omega}})
\label{eq:barX'}
\eeq 
and hence:
\begin{equation}
    \bar{X}'(\mu_\theta^+) \leq \bar{\Gamma}'(\mu_\theta^+) < 0 \ .
    \label{eq:61}
\end{equation}
Combining Eqs.~\eqref{eq:population},~\eqref{eq:58}, and~\eqref{eq:app_dFg_x}, we thus find the following equivalent sufficient condition for $\pdv{\rho_{00}}{s_p} \Big\vert_{s_p = \mu^+_\theta} > 0$:
\begin{align}
    \bar{\Gamma}'\pqty{\mu_\theta^+}\int_0^1e^{\bar{S}x}\dd{x} &>s_d \bar{\Gamma}\pqty{\mu_\theta^+}\bar{X}'\pqty{\mu_\theta^+}\int_0^1e^{\bar{S}x}\pqty{1-x}\dd{x} \notag\\
    & \quad +P_\varphi \bar{X}'\pqty{\mu_\theta^+} \ .\label{eq:mu_theta_condition_1}
\end{align}
Then, by explicitly carrying out the integrals, replacing one factor of $\bar{S}$ by $s_d\bar{X}\pqty{\mu_\theta^+}$, and dividing inequality~\eqref{eq:mu_theta_condition_1} by $\bar{X}'\pqty{\mu_\theta^+}<0$, we have:
\begin{equation}\label{eq:mu_theta_condition_2}
    \frac{\bar{\Gamma}'\pqty{\mu_\theta^+}}{\bar{X}'\pqty{\mu_\theta^+}}\frac{e^{\bar{S}}-1}{\bar{S}} < \frac{\bar{\Gamma}\pqty{\mu_\theta^+}}{\bar{X}\pqty{\mu_\theta^+}}\bigg(\frac{e^{\bar{S}}-1}{\bar{S}}-1\bigg) + P_\varphi \ .
\end{equation}

Let us denote
\bes
\begin{align}
\label{eq:Pbar}
    \bar{P}_{\text{th}}\pqty{s} &= \frac{\bar{\Gamma}\pqty{s}}{\bar{X}\pqty{s}} = \frac{1}{1+e^{-\beta\bar{\Omega}\pqty{s}}} \\
    \bar{Q}\pqty{s} &= \frac{\bar{\Gamma}'\pqty{s}}{\bar{X}'\pqty{\mu_\theta^+}\bar{P}_{\text{th}}\pqty{s}} \ .
\end{align}
\ees
We then note that 
\begin{equation}
    1-\bar{Q}\pqty{\mu_\theta^+} = \frac{\beta\bar{\Omega}'e^{-\beta\bar{\Omega}}\bar{\Gamma}}{\beta\bar{\Omega}'e^{-\beta\bar{\Omega}}\bar{\Gamma}-\pqty{1+e^{-\beta\bar{\Omega}}}\bar{\Gamma}'} > 0 \ ,
\end{equation}
where the inequality holds at $\mu_\theta^+$: we know the gap is increasing so $\bar{\Omega}'\pqty{\mu_\theta^+}>0$ and the numerator is positive, and $\bar{\Gamma}'\pqty{\mu_\theta^+}<0$ by Assumption~\ref{as:1}, so the denominator is also positive.

Eq.~\eqref{eq:mu_theta_condition_2} can thus be rewritten as
\begin{equation}
\label{eq:mu_theta_condition_3}
    \frac{e^{\bar{S}}-1}{\bar{S}} > \frac{1-P_\varphi / \bar{P}_{\text{th}}\pqty{\mu_\theta^+}}{1-Q\pqty{\mu_\theta^+}}  \equiv x \ .
\end{equation}
Note that the function $\frac{e^{\bar{S}}-1}{\bar{S}}\geq 1$ and is monotonically increasing for $\bar{S}\geq 0$. Therefore the inequality is automatically satisfied for $x<1$ by any $\bar{S}>0$.

Let us denote by $\bar{S}^*$ the solution of inequality~\eqref{eq:mu_theta_condition_3} replaced by an equality; this transcendental equation has a formal solution in terms of the Lambert-$W$ function~\cite{Lambert-W}, i.e. the inverse function of $f(W) = We^W$: 
\beq
\bar{S}^*(x) = -\frac{1}{x}\bigg(1 + x W_{-1}(-e^{-1/x}/x) \bigg)  = s_d^*\bar{X}\pqty{\mu_\theta^+} \ ,
\label{eq:Lambert}
\eeq
where $W_{-1}(z)$ is one of the two real branches of $W(z)$ satisfying 
\begin{equation}\label{eq:W_minus_branch}
    W_{-1}(z) \leq -1 \quad -1/e \leq z <0 \ ,
\end{equation}
with $W_{-1}(-1/e) = -1$. The function $\frac{e^{\bar{S}}-1}{\bar{S}}$ is monotonically increasing, so  inequality~\eqref{eq:mu_theta_condition_3} is satisfied for all $\bar{S}  > \max(0,\bar{S}^*)$, i.e., for all $s_d > s_d^* = \max[0,\bar{S}^*/\bar{X}\pqty{\mu_\theta^+}]$. 

We can therefore replace condition \eqref{eq:mu_theta_condition_3} with
\begin{myassumption}
\label{as:2}
\beq
s_d \bar{\Gamma}\pqty{\mu_\theta^+} > \bar{S}^*(x)\bar{P}_{\text{th}}\pqty{\mu_\theta^+} \ ,
\label{eq:as:2}
\eeq
where $\bar{P}_{\text{th}}\pqty{\mu_\theta^+}$ is the thermal ground state probability at $s=\mu_\theta^+$.
\end{myassumption}
Moreover, using the recursive expression
\begin{equation}
    W_{-1}(z) = \ln(-z) - \ln(-W_{-1}(z)) \ ,
\end{equation}
Eq.~\eqref{eq:Lambert} can be written as
\begin{equation}
    \bar{S}^*(x) = \ln{x} + \ln(-W_{-1}(-e^{-1/x}/x)) < 2\ln(x) \ . 
    \label{eq:barSbound}
\end{equation}
The above upper bound is derived in Appendix~\ref{app:barSbound}. 

Thus we can replace Eq.~\eqref{eq:as:2} by $s_d \bar{\Gamma}\pqty{\mu_\theta^+} > 2\ln(x)$, which grows very mildly, and can for practical purposes be replaced by an $O(1)$ constant. This is how Assumption~\ref{as:2} is stated in Theorem~\ref{thm:opt-pausing}.
  
Note that the case $s_d^* = 0$ arises only when $x<0$ in Eq.~\eqref{eq:mu_theta_condition_3} [since then the solution $\bar{S}^*$ given by Eq.~\eqref{eq:Lambert} is negative]. This, in turn arises when $P_\varphi >  \bar{P}_{\text{th}}(\mu_\theta^+)$, i.e., when the instantaneous ground state probability is greater than the thermal ground state probability, both at $\tau=\mu_\theta^+$. Indeed, this conforms with the expectation this in this case pausing is not advantageous.

\subsubsection{Proof of $ \pdv{\rho_{00}}{s_p} \Big\vert_{s_p = 1} = \pdv{\mathcal{F}_g}{s_p}\Big\vert_{s_p = 1} + \pdv{\mathcal{F}_d}{s_p} \Big\vert_{s_p = 1} < 0$}

Note that using Eq.~\eqref{eq:barX'} we have 
$\bar{X}'(1) < 0$ since, by Assumption~\ref{as:1}, ${\bar{\Gamma}}'(1)<0$, and $\bar{\Omega}'(1)>0$ since the gap grows at the end of the anneal, as per Eq.~\eqref{eq:gaussian_gap} (this need not always be the case~\cite{Altshuler2010}).

Combining Eqs.~\eqref{eq:app_dFd_1} and~\eqref{eq:app_dFg_4}, we thus have%
\footnote{
Note that if $\bar{X}'(1) > 0$, the RHS of Eq.~\eqref{eq:app_sp_1} is automatically negative, since the prefactor $\abs{\bar{X}'(1)}$ comes from $-\bar{X}'(1)$, and the $-\abs{\bar{X}'(1)}$ inside the square brackets becomes $\bar{X}'(1)$, so every term inside these brackets is positive. In this case Assumptions 3 and 4 of Theorem~\ref{thm:opt-pausing} can be dropped. However, $\bar{X}'(1) < 0$ in our model.}
\begin{align}
 \label{eq:app_sp_1} 
 &    \pdv{\mathcal{F}_g}{s_p} \bigg\vert_{s_p=1} + \pdv{\mathcal{F}_d}{s_p} \bigg\vert_{s_p=1} = s_d|\bar{X}'(1)|\bigg[P_\varphi  G(\mu_\theta^+)  + \notag
\\
&\quad  \int_{\mu^+_\theta}^{1}\dd{\tau}G\pqty{\tau}\Gamma\pqty{\tau} 
-\bigg\vert\frac{\bar{\Gamma}'(1)}{s_d \bar{X}'(1)}\bigg\vert  \int_{1}^{\tau_f}\dd{\tau}G\pqty{\tau} \notag \\
&\quad + \bar{\Gamma}(1)\int_{1}^{\tau_f}\dd{\tau} G(\tau)\pqty{1+\frac{1}{s_d}(1-\tau)} \bigg] \ .
\end{align}
Therefore it suffices to find a condition under which the expression inside the square brackets in Eq.~\eqref{eq:app_sp_1} is negative. Our strategy for doing so is to replace this expression with a simpler but negative upper bound, and iterating this until we arrive at a conceptually simple final expression.

Now note that for all $\tau \in \bqty{1, \tau_f}$:
\begin{align}
\label{eq:Gineq}
G(\tau) &= \exp[-\int_{\tau}^{\tau_f}\dd{\tau'}X(\tau')] = \exp[-(\tau_f-\tau)\bar{X}(1)] \ ,
\end{align}
since when $s_p=1$ all the schedule-dependent functions are constant for $\tau$ in the range $[1,\tau_f]$, as a result of Eq.~\eqref{eq:cases}.

Therefore,
 \begin{align}
     \int_{1}^{\tau_f}\dd{\tau}G(\tau) &= 
      \frac{1-e^{-s_d\bar{X}(1)}}{\bar{X}(1)} \ ,
 \end{align}
and
\bes
\begin{align}
& \int_{1}^{\tau_f}\dd{\tau} G(\tau)\pqty{1+\frac{1}{s_d}(1-\tau)} \\
&\quad = \frac{1-e^{-s_d\bar{X}(1)}[1+s_d\bar{X}(1)]}{s_d\bar{X}(1)^2} \ .
\end{align}
\ees

We can find upper bounds involving $G$ by using
\begin{equation}\label{eq:XgeGamma}
    X(\tau) = (1+e^{-\beta\Omega(\tau)})\Gamma(\tau) \ge \Gamma(\tau)\ .
\end{equation}
Thus, using Eqs.~\eqref{eq:Gineq} and~\eqref{eq:XgeGamma} we have: 
\bes
\begin{align}
&    \int_{\mu^+_\theta}^{1}\dd{\tau}G\pqty{\tau}\Gamma\pqty{\tau} =e^{-s_d \bar{X}\pqty{1}}\int_{\mu^+_\theta}^{1}\dd{s} e^{-\int_{\tau}^{1}\dd{\tau'}X(\tau')}\bar{\Gamma}(s) \\
    &\qquad \le e^{-s_d \bar{X}\pqty{1}}\int_{\mu^+_\theta}^{1}\dd{s} e^{-\int_{s}^{1}\dd{s'}\bar{\Gamma}(s')}\bar{\Gamma}(s) \\
    & \qquad= e^{-s_d \bar{X}\pqty{1}} \pqty{1 - e^{-\int_{\mu_\theta^+}^{1}\bar{\Gamma}\pqty{s}\dd{s}}} \ ,
\end{align}
\ees
and 
\bes
\begin{align}
G(\mu_\theta^+) &
= \exp[-\int_{\mu_\theta^+}^{1}\dd{\tau}X(\tau)]\exp[-\int_{1}^{1+s_d}\dd{\tau}X(\tau)] \\
& \leq \exp[-\int_{\mu_\theta^+}^{1}\dd{s}\bar{\Gamma}(s)]\exp[-s_d\bar{X}(1)] \ .
\end{align}
\ees

Next, let us rewrite $\vert\frac{\bar{\Gamma}'(1)}{ \bar{X}'(1)}\vert$ by using Eq.~\eqref{eq:barX'}. First, let 
\begin{equation}
\label{eq:epsilon}
\epsilon \equiv \frac{\bar{\Gamma}\pqty{1}\beta \bar{\Omega}'(1)}{|\bar{\Gamma}'(1)|(1+e^{\beta\bar{\Omega}(1)})} \ .
\end{equation}
In Appendix~\ref{app:proof_epsilon}, we argue that for spectral densities with an exponential tail (e.g., the Ohmic case we consider) $\epsilon \ll 1$ for sufficiently large $\beta\bar{\Omega}(1)$. 

Then, using Assumption~\ref{as:1} again:
\bes
\label{eq:70}
\begin{align}
\bigg\vert\frac{\bar{\Gamma}'(1)}{ \bar{X}'(1)}\bigg\vert &= \frac{|\bar{\Gamma}'|}{\vert {\bar{\Gamma}}'(1+e^{-\beta \bar{\Omega}})-\beta \bar{\Omega}' e^{-\beta \bar{\Omega}}\bar{\Gamma} \vert} \\
&= \frac{1}{(1+\epsilon)(1+e^{-\beta\bar{\Omega}(1)})} = \frac{1}{1+\epsilon}\bar{P}_{\text{th}}(1)\ ,
\end{align}
\ees
    
Defining 
\beq
\lambda \equiv s_d\bar{X}(1) = s_d(1+e^{-\beta\bar{\Omega}(s)})\bar{\Gamma}(s)\ ,
\eeq 
we can now combine all these bounds to provide an upper bound on the expression in square brackets in Eq.~\eqref{eq:app_sp_1}:
\begin{align}
&\bigg[ \cdots \bigg] \leq P_\varphi  e^{-\lambda}e^{-\int_{\mu_\theta^+}^{1}\bar{\Gamma}\pqty{s}\dd{s}}\notag  \\
&\quad  + e^{-\lambda} - e^{-\lambda} e^{-\int_{\mu_\theta^+}^{1}\bar{\Gamma}\pqty{s}\dd{s}}  
-\frac{1}{1+\epsilon}\bar{P}_{\text{th}}(1) 
\frac{1-e^{-\lambda}}{\lambda}\notag \\
&\quad + \bar{P}_{\text{th}}(1)\frac{1-e^{-\lambda}(1+\lambda)}{\lambda} \ ,
\label{eq:66}
\end{align}
an expression we require to be negative. We thus arrive at the sufficient condition
\begin{equation}
\int_{\mu_\theta^+}^{1}\bar{\Gamma}\pqty{s}\dd{s}   \leq \ln \bigg( \frac{1-P_\varphi }{1-\bar{P}_{\text{th}}(1)F\pqty{\lambda}} \bigg) \ ,
\label{eq:sc4}
\end{equation}
where
\begin{equation}\label{eq:FLambda}
    F\pqty{\lambda} = 1-\frac{e^{\lambda}-1}{\lambda}\frac{\epsilon}{1+\epsilon} \ .
\end{equation}
Since $\bar{\Gamma}\pqty{s}\geq 0$ for all $s$, the bound must positive to be sensible. Thus the argument of the logarithm must be lower bounded by $1$. In order for the bound in Eq.~\eqref{eq:sc4} to be positive it is therefore sufficient to require that
\beq
\label{eq:fcondition}
P_\varphi / \bar{P}_{\text{th}}(1) <  F\pqty{\lambda} \ ,
\eeq
Without loss of generality, a sufficient condition for Eq.~\eqref{eq:fcondition} is: $\exists c > 1$ such that
\begin{equation}\label{eq:fbound}
    F\pqty{\lambda} > \frac{1}{1+c\epsilon} > \frac{P_\varphi}{\bar{P}_{\text{th}}(1)} \ .
\end{equation}
which is not unreasonable because in practice, we would expect $\bar{P}_{\text{th}}(1) \sim 1$, $P_\varphi \lesssim 0.5$ (for a hard instance) and $\epsilon \ll 1$. Substituting Eq.~\eqref{eq:FLambda} into Eq.~\eqref{eq:fbound}, we have
\begin{equation}
    \frac{e^{\lambda}-1}{\lambda} < 1 + \frac{c-1}{1+c\epsilon} \equiv x \ .
    \label{eq:80}
\end{equation}
Recall that the solution $\lambda^*(x)$ to this inequality considered as an equality is the Lambert-$W$ function [Eq.~\eqref{eq:Lambert}, with $\lambda^*$ replacing $\bar{S}^*$], and the function $ \frac{e^{\lambda}-1}{\lambda}$ is monotonically increasing. Therefore Eq.~\eqref{eq:80} is satisfied as long as $\lambda \equiv s_d\bar{X}(1) < \lambda^*$:
\begin{myassumption}
\label{as:3}
\begin{equation}
    s_d \bar{\Gamma}\pqty{1} < \lambda^*(x) < 2\ln(x) \ ,
\end{equation}
\end{myassumption}
where as before we may view $\lambda^*$ in practice as an $O(1)$ constant. 
This is how Assumption~\ref{as:3} is stated in Theorem~\ref{thm:opt-pausing}.

Finally, using the lower bound~\eqref{eq:fbound}, Eq.~\eqref{eq:sc4} can be replaced with:
\begin{align}
& \int_{\mu_\theta^+}^{1}\bar{\Gamma}\pqty{s}\dd{s}   \leq 
\ln\bigg(\frac{1-P_\varphi}{1-\bar{P}_{\text{th}}(1)/(1+c\epsilon)}\bigg) \ .
\label{eq:sc5}
\end{align}
Rewritten as
\begin{myassumption}
\label{as:4}
\beq
1-(1-P_\varphi)e^{-\int_{\mu_\theta^+}^{1}\bar{\Gamma}\pqty{s}\dd{s}} \leq \frac{1}{1+c\epsilon}\bar{P}_{\text{th}}(1) \leq \bar{P}_{\text{th}}(1)\ ,
\label{eq:as4}
\eeq
\end{myassumption}
this can be interpreted as follows: the excited state population right after crossing the minimum gap is $1-P_\varphi$, and after multiplying this by $\exp[-\int_{\mu_\theta^+}^{1}\bar{\Gamma}\pqty{s}\dd{s}]$ we have what is left of this excited state population at the end of the anneal. On the other hand, $\bar{P}_{\text{th}}(1)$ is the thermal ground state population assuming equilibration. In other words, Eq.~\eqref{eq:as4} states that the actual ground state population reached at the end of the anneal is less than the thermal ground state population, i.e., the system has not fully equilibrated. This is the version of Assumption~\ref{as:4} given in Theorem~\ref{thm:opt-pausing}.

\section{Conclusions}
\label{sec:conc}
We have established numerically as well as analytically, via an open system analysis of two-level system models, that pausing-induced quantum thermal relaxation can play a positive role in quantum annealing, at least according to the success probability metric. More specifically, we have shown here that under certain conditions on the relaxation rate after the minimum gap is crossed, the ground state probability increases when pausing occurs before the end of the anneal. For this to occur the relaxation rate should be decreasing both after the minimum gap is crossed and at the end of the anneal, and cumulatively small over this interval, so that the system does not fully thermally equilibrate. In addition, the pause duration should be large relative to the inverse relaxation rate after the minimum gap is crossed, but small relative to the inverse relaxation rate at the end of the anneal. This provides a set of sufficient conditions relating to non-equilibrium dynamics and incomplete quantum thermal relaxation that explain the improved pause-based performance reported in a series of recent experimental quantum annealing studies~\cite{marshall_power_2019,venturelli_reverse_2019,vinci2019path}. The framework we have established also provides tools to solve for the optimal pause position $s_p^*$ [Eq.~\eqref{eq:app_derivative_0}]. We expect that analytic solutions for $s_p^*$ can be derived in a problem-specific manner with further approximations.

Our results leave open a number of interesting questions for future studies. We have not determined the optimal pause time, nor did we demonstrate that pausing guarantees a quantum speedup. Indeed, computationally meaningful metrics such as the time-to-solution~\cite{speedup} may not be enhanced due to extra time cost incurred due to pausing~\cite{izquierdo_ferromagnetically_2020}, and it is also possible that classical models of quantum annealing, such as the spin-vector Monte Carlo algorithm~\cite{SSSV}, similarly benefit from pausing~\cite{Marshall-private}. 

\acknowledgments
The authors are grateful to Jenia Mozgunov and Humberto Munoz Bauza for useful discussions and feedback. We used the Julia programming~\cite{bezanson_julia:_2017} and the DifferentialEquations.jl package~\cite{rackauckas_differentialequations.jl_2017} for all our numerical calculations. 

The research is based upon work (partially) supported by the Office of
the Director of National Intelligence (ODNI), Intelligence Advanced
Research Projects Activity (IARPA) and the Defense Advanced Research Projects Agency (DARPA), via the U.S. Army Research Office
contract W911NF-17-C-0050. The views and conclusions contained herein are
those of the authors and should not be interpreted as necessarily
representing the official policies or endorsements, either expressed or
implied, of the ODNI, IARPA, DARPA, or the U.S. Government. The U.S. Government
is authorized to reproduce and distribute reprints for Governmental
purposes notwithstanding any copyright annotation thereon.

\FloatBarrier

\appendix

\section{Single-qubit adiabatic frame}
\label{app:single_qubit_adiabatic_frame}

We recall how to transform the Hamiltonian
    $H_{\mathrm{S}}(s) = -\frac{1}{2} \big[ A(s)Z +  B(s)X \big]$
into the adiabatic frame. First, we reparametrize the annealing schedules in terms of the gap $\Omega(s)$ and rotation angle $\theta(s)$:
\begin{equation}
    A(s) = \Omega(s)\cos\theta(s),\quad
    B(s) = \Omega(s)\sin\theta(s)\ .
\end{equation}
Then, we rescale it to a dimensionless quantity by a change of variable $s = t / t_f$ in Von Neumann equation
\begin{align}
    \dv{\rho}{s} = -i\comm{t_f H_\mathrm{S}(s)}{\rho} \ .
    \label{eq:vN}
\end{align}
Finally, we rotate the system with respect to the unitary
\begin{equation}
    U = \exp[i \theta(s) Y / 2] \ .
\end{equation}
The dimensionless interaction picture Hamiltonian is
\begin{align}
    \tilde{H}_\mathrm{S}(s) &= t_fU^\dagger(s) H_\mathrm{S}(s) U(s) -i U^\dagger(s) \pdv{}{s}U(s)\nonumber \\
    &= \frac{1}{2}\pqty{\dv{\theta}{s}Y-t_f\Omega\pqty{s}Z} \ .
\end{align}
The important observation is that in this rotating frame the eigenstates of the $Z$ operator always align with the instantaneous energy eigenstates of the original Hamiltonian. So, it can be thought of as a co-rotating frame of the adiabatic basis. This co-rotating frame, which we refer to as the adiabatic frame throughout this paper, can be extended to the multi-qubit case as described next.

\section{Multi-qubit adiabatic frame}
\label{app:multi_qubit_adiabatic_frame}

Starting from the von Neumann equation~\eqref{eq:vN},
we have
\begin{align}
    &\sum_{nm} \dot{\rho}_{nm}\dyad{n}{m} + \rho_{nm}\dyad{\dot{n}}{m} + \rho_{nm}\dyad{n}{\dot{m}} = \nonumber \\
    & \quad -it_f\sum_{nm}\pqty{E_n-E_m}\rho_{nm}\dyad{n}{m} \ . \label{eq:qom_rho}
\end{align}
To derive an effective equation of motion for the density matrix in the adiabatic frame $\tilde{\rho} = \bqty{\rho_{nm}}$, we wish to write $\dyad{\dot{n}}{m}$ and $\dyad{n}{\dot{m}}$ in $\Bqty{\ket{n}}$ basis. For example, the first term can be written as
\begin{equation}
\label{eq:dot_dyad}
    \dyad{\dot{n}}{m} = \sum_{n'}\braket{n'}{\dot{n}}\dyad{n'}{m} \ .
\end{equation}
It is important to emphasize that $\ket{\dot{n}}$ means the derivative with respect to $s$ of the $n$'th eigenstate of an $s$-dependent Hamiltonian and does not obey the Schr\"odinger equation. We assume that $H_\mathrm{S}(s)$ is non-degenerate.
We show below that an explicit formula for $\braket{m}{\dot{n}}$ is given by 
\begin{equation}
\label{eq:gap_geometric_term}
    \langle m | \dot{n}\rangle= \frac{\left\langle m(s)\left|\frac{\mathrm{d} H_\mathrm{S}(s)}{\mathrm{d} s}\right| n(s)\right\rangle}{E_{n}(s)-E_{m}(s)} \delta_{mn} \ ,
\end{equation}
which directly leads to $\braket{m}{\dot{n}} = - \braket{n}{\dot{m}}^*$. Substituting Eq.~\eqref{eq:dot_dyad} into Eq.~\eqref{eq:qom_rho}, we obtain
\begin{align}
    \dot{\rho}_{nm} &= -it_f\pqty{E_n-E_m}\rho_{nm} \nonumber \\
    & \quad -\sum_{n'\neq n}\rho_{n'm}\braket{n}{\dot{n'}} - \sum_{m'\neq m}\rho_{nm'}\braket{\dot{m'}}{m} \nonumber \\
    & = -it_f\pqty{E_n-E_m}\rho_{nm} \nonumber \\
    &\quad -i\bqty{-i\sum_{n'\neq n}\braket{n}{\dot{n'}}\rho_{n'm} +i \sum_{m'\neq m}\rho_{nm'}\braket{m'}{\dot{m}}}
\end{align}
for each $\rho_{nm}$. Thus, an effective Hamiltonian satisfying $\dot{\tilde{\rho}} = -i\comm{\tilde{H}}{\tilde{\rho}}$ for $\tilde{\rho} = \bqty{\rho_{nm}}$ is the one given in Eq.~\eqref{eq:effective_H} in the main text (where we assumed that $H_\mathrm{S}(s)$ is real).

Let us now prove Eq.~\eqref{eq:gap_geometric_term}.
%
Writing the system Hamiltonian in its eigenbasis as
\begin{equation}
    H_\mathrm{S}(s) = \sum_n E_n\pqty{s}\dyad{n} \ ,
\end{equation}
we will derive the expression for the geometric term $\braket{m}{\dot{n}}$ under the assumption that $H_\mathrm{S}(s)$ is non-degenerate and real. Taking the derivative of $H_\mathrm{S}\ket{n} = E_n \ket{n}$ with respect to $s$ and multiplying both sides by $\bra{m}$, we have
\begin{equation}\label{eq:app_geo_exp}
    \matrixel{m}{\dot{H}_\mathrm{S}}{n} - \dot{E}_n\braket{m}{n} = \pqty{E_n-E_m}\braket{m}{\dot{n}} \ .
\end{equation}
If $m \neq n$ and $\ket{m}$, $\ket{n}$ are non-degenerate, the above expression reduces to Eq.~\eqref{eq:gap_geometric_term}. For the case where $m=n$, Eq.~\eqref{eq:app_geo_exp} becomes
\begin{equation}\label{eq:app_nHn_E}
    \matrixel{n}{\dot{H}_\mathrm{S}}{n} = \dot{E}_n \ .
\end{equation}
On the other hand, we can also take the derivative of $\matrixel{n}{H_\mathrm{S}}{n} = E_n$, which leads to
\begin{equation}\label{eq:app_cmn}
    \matrixel{\dot{n}}{H_\mathrm{S}}{n} + \matrixel{n}{\dot{H}_\mathrm{S}}{n} + \matrixel{n}{H_\mathrm{S}}{\dot{n}} = \dot{E}_n \ .
\end{equation}
By cancelling out $\matrixel{n}{\dot{H}_\mathrm{S}}{n}$ and $\dot{E}_n$ and noticing that $\braket{n}{\dot{n}}$ is real, we can deduce that $\braket{n}{\dot{n}}=0$. It is important to note that the eigenvector $\ket{n}$ is only uniquely determined up to a constant factor of $\pm 1$. As a result, we need to implement a continuous constraint 
\begin{equation}
\label{eq:app_continuous_condition}
    \lim_{\Delta s \to 0 }\braket{n\pqty{s}}{n\pqty{s+\Delta s}} = 1
\end{equation}
to ensure the continuity of the geometric term.

This result can be extended to a general complex-valued Hamiltonian. In this case, the eigenvectors of the Hamiltonian are uniquely determined up to a constant of unit modulus. However, by enforcing the continuity condition~\eqref{eq:app_continuous_condition}, we have an analytic $\ket{n(s)}$ that also satisfies $\braket{n}{\dot{n}} = 0$.

A method to calculate $\braket{m}{\dot{n}}$ for degenerate Hamiltonian is provided in Ref.~\cite{andrew_computation_1998}. A special case that is not discussed in this reference is when two or more states $\ket{m}$ and $\ket{n}$ become degenerate in a closed interval $s \in \bqty{s_a, s_b}$. In such a case, we can still obtain a pair of orthogonal states $\ket{m(s_a)}$ and $\ket{n(s_a)}$ by enforcing the continuity condition \eqref{eq:app_continuous_condition} across the boundary. Within the interval $\braket{m}{\dot{n}}$s are usually $0$ in practice. A sufficient condition for this is
\begin{equation}
\label{eq:app_degenerate_condition}
    \lim_{\Delta s \to 0} \braket{m\pqty{s}}{n\pqty{s+\Delta s}} /\Delta s = 0
\end{equation}
for all $s \in [s_a, s_b] $. By expanding $\ket{n(s+\Delta s)}$ as a Taylor series in $\Delta s$, Eq.~\eqref{eq:app_degenerate_condition} reduces to
\begin{equation}
    \lim_{\Delta s \to 0} \braket{m}{n} / \Delta s + \braket{m}{\dot{n}} + O(\Delta s) = 0 \ ,
\end{equation}
which implies $\braket{m}{\dot{n}}=0$. One example of condition \eqref{eq:app_degenerate_condition} is when the transverse field becomes zero during the anneal and the problem Hamiltonian has degenerate excited states.

\section{Annealing angle behavior for previously studied small gap quantum annealing examples}
\label{app:multi_qubits_example}

Here we provide a brief look at the annealing angle aspect of two previously studied quantum annealing examples with small gaps. In the following examples, all the results are produced with a linear schedule instead of the D-Wave schedule used in the references. 

\begin{figure}[t]
    \centering
    ~ 
        \subfigure[\ ]{\includegraphics[width=.4\textwidth]{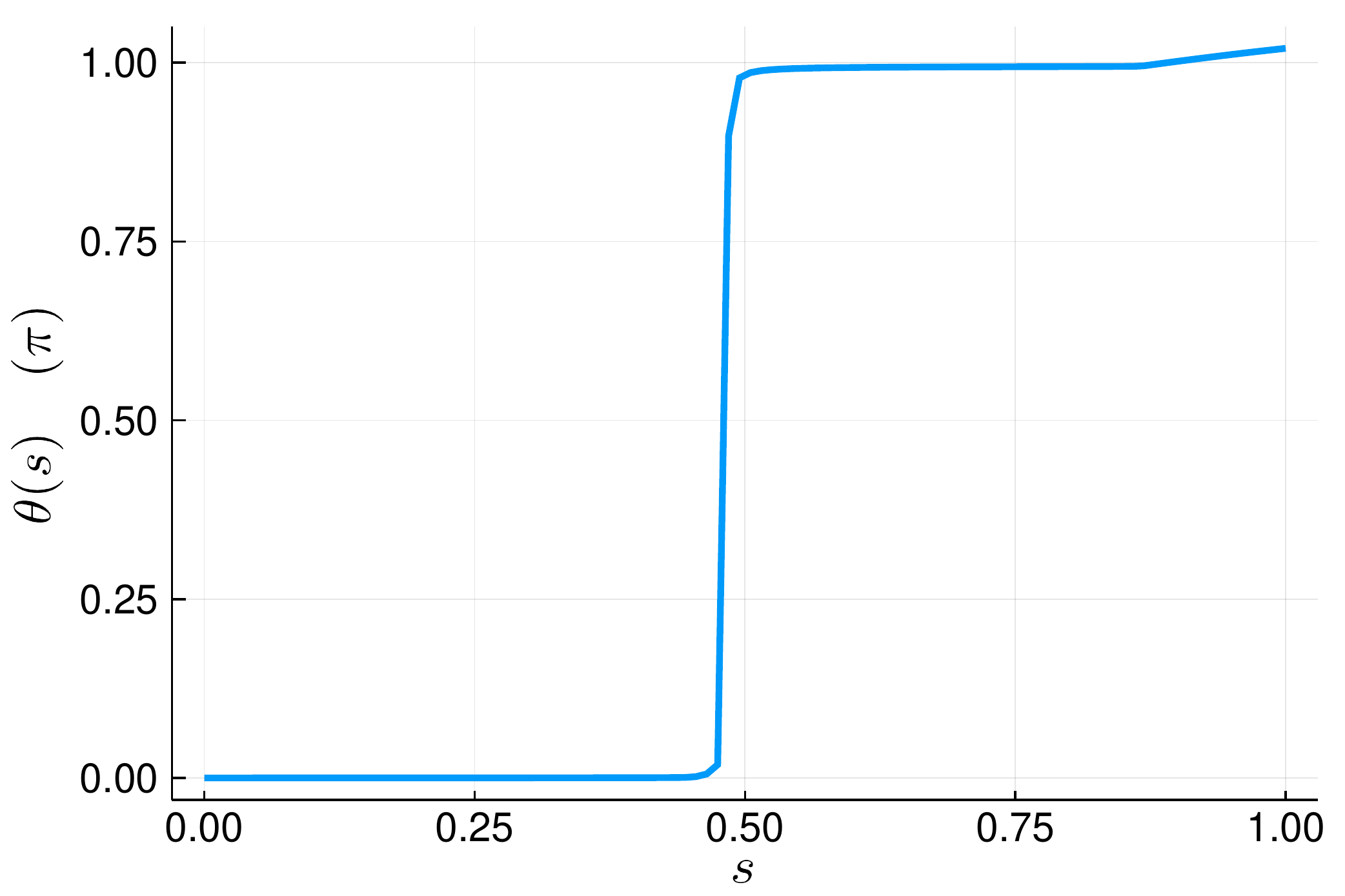}}
        \subfigure[\ ]{\includegraphics[width=.4\textwidth]{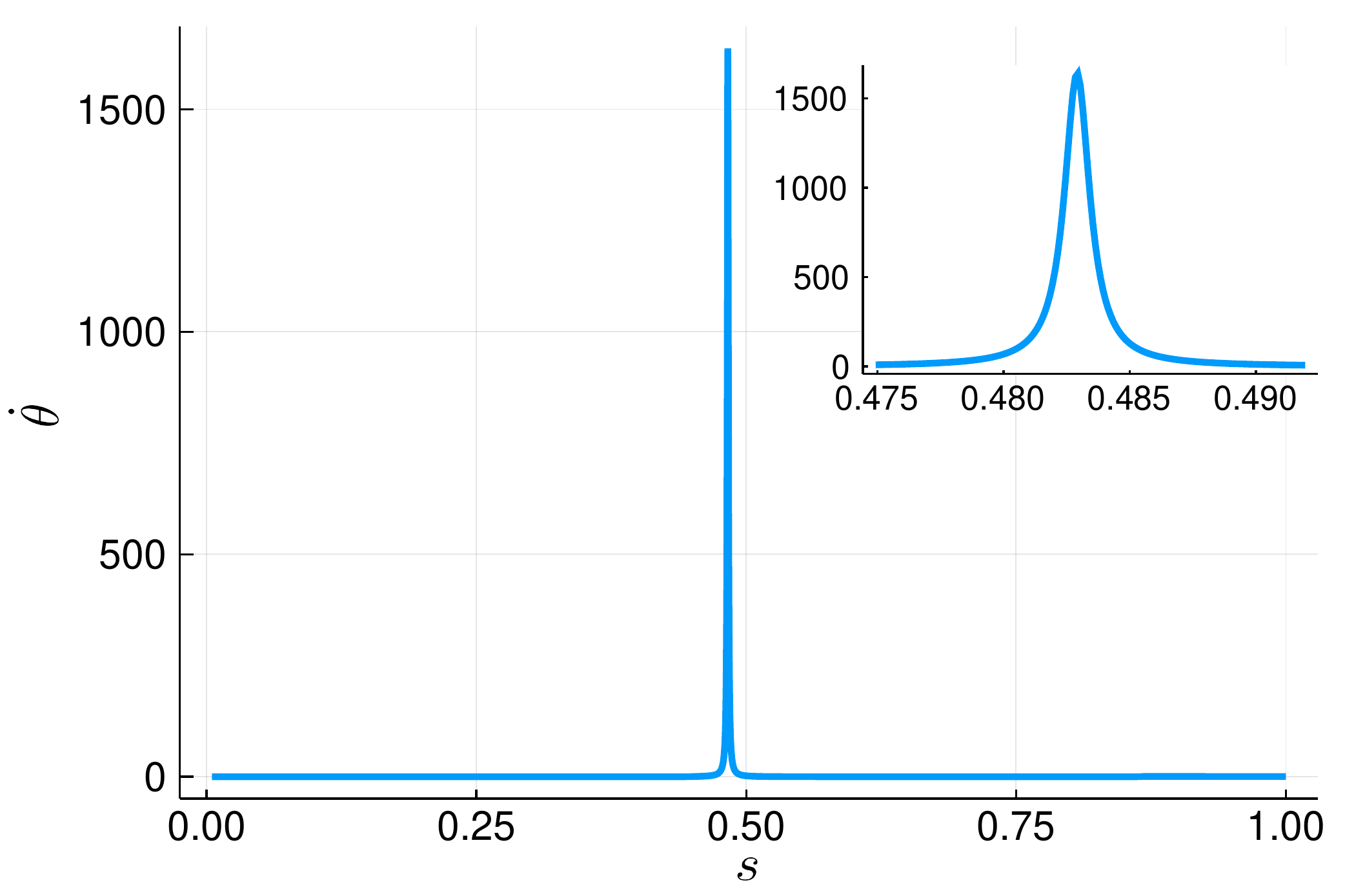}}
\subfigure[\ ]{\includegraphics[width=.4\textwidth]{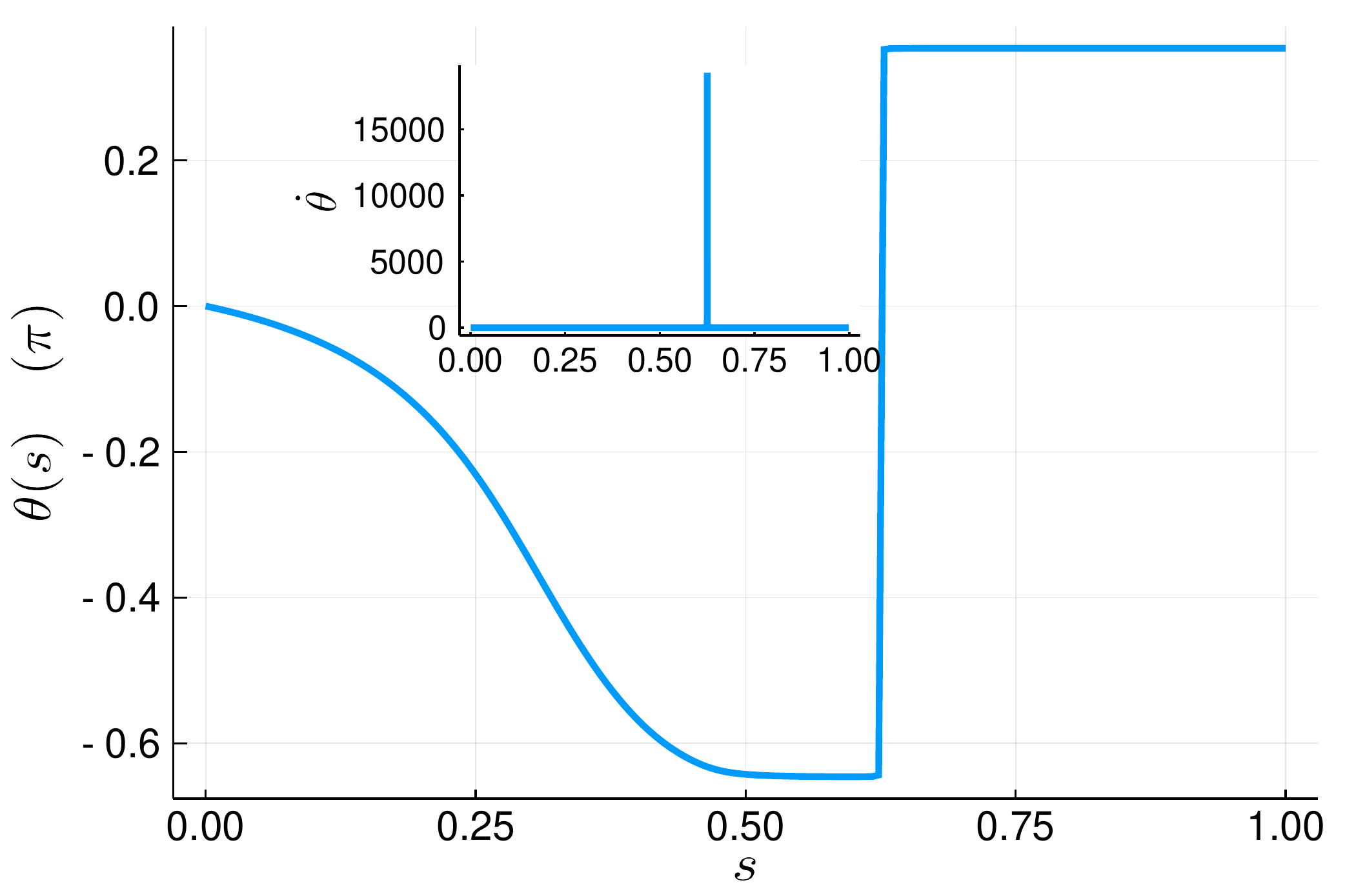}}
\caption{(a) Annealing angle for the $p$-spin model with $n=20$, $p=19$~\cite{passarelli_improving_2019}. The inset in zooms in around the peak of angular progression, shown in (b). In (c) we show the annealing angle for the D-Wave $16$-qubit gadget~\cite{dickson_thermally_2013}. The inset illustrates the angular progression.}
\label{fig:p_spin_dtheta}
\end{figure}

The first example, shown in Fig.~\ref{fig:p_spin_dtheta}(a) and (b), is the $p$-spin model~\cite{passarelli_improving_2019}. The angular progression is localized around $s=0.483$. Across the region, there is a $\pi$ jump of the annealing angle. 

The second example, shown in Fig.~\ref{fig:p_spin_dtheta}(c), is the D-Wave $16$-qubit gadget problem~\cite{dickson_thermally_2013}. Unlike the $p$-spin model, the annealing angle in this case does not stay zero during the first half of the anneal. There is still a sharp $\pi$ jump across the minimum gap region.


\section{Easy problem with small gap: constant angular progression}
\label{app:const_dtheta}

\begin{figure}[htb]
\includegraphics[width=\columnwidth]{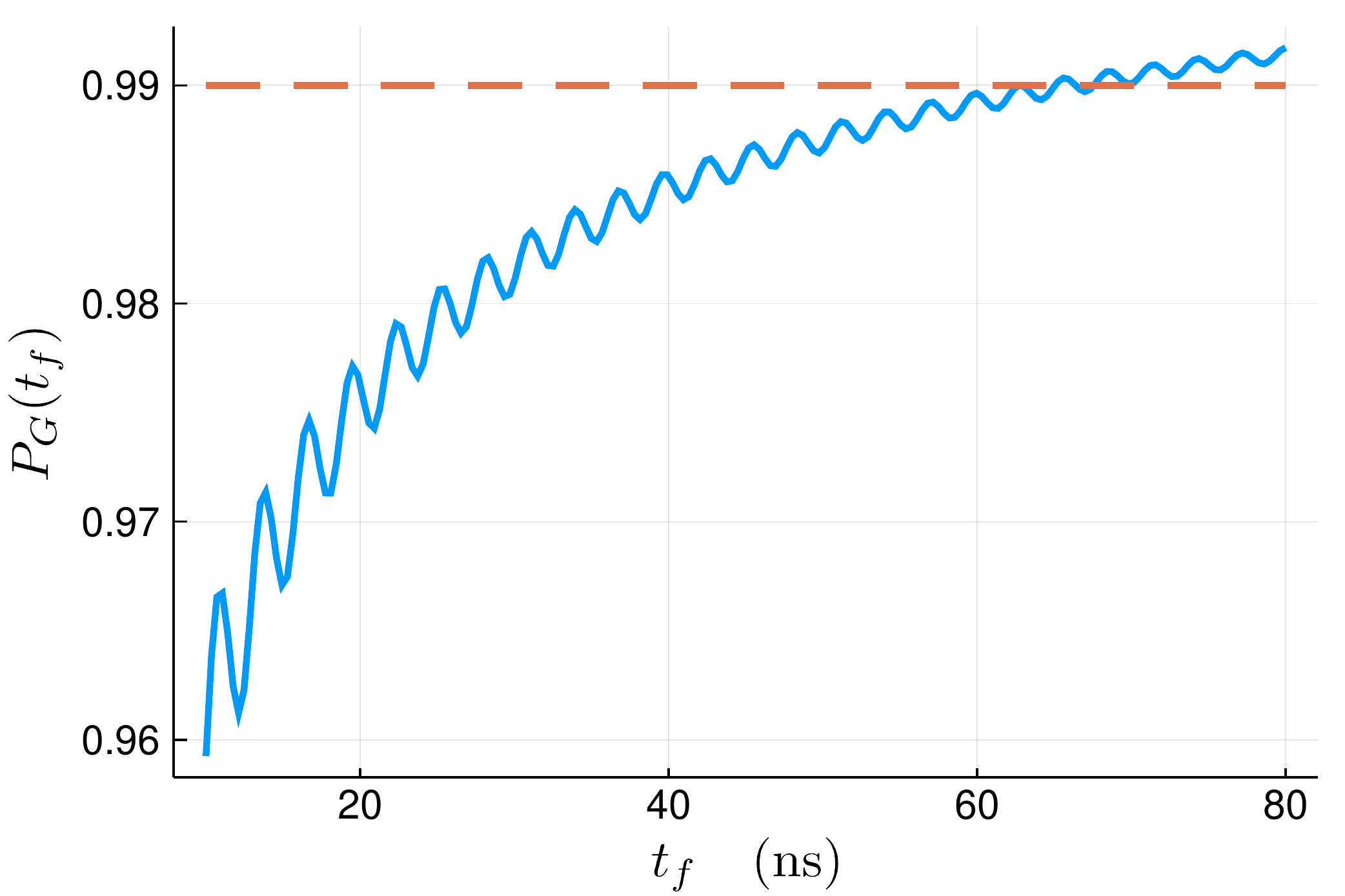}
\caption{Closed system success probability versus total annealing time. The gap is chosen according to Eq.~\eqref{eq:gaussian_gap} with parameters $\mu_g=0.5$, $\alpha_g = 0.5$, $E_0=15/\pi$ GHz and $\Delta=0.001$.
}
\label{fig:const_angular_momentum}
\end{figure}

The simplest example we can construct of an easy problem with a small gap is one with the same small gap structure as in Eq.~\eqref{eq:gaussian_gap} but a constant $\dot{\theta}\pqty{s} = \pi / 2$. If we define the adiabatic time scale $t_{\mathrm{ad}}$ as the point where
\begin{equation}
    P_G(t) \ge 0.99 \quad \forall t \ge t_{\mathrm{ad}} \ ,
\end{equation}
we find in the numerical example shown in Fig.~\ref{fig:const_angular_momentum}, that the adiabatic time scale is much smaller than the inverse gap
\begin{equation}
    t_{\mathrm{ad}} \ll h / (E_0 \Delta)^2 \approx 142502 (\mathrm{ns})\ ,
\end{equation}
where
\begin{equation}
    h = \max_{s\in[0,1]}\norm{\dot{H}(s)} \ ,
\end{equation}
and $\norm{\cdot}$ is the operator norm [for the Frobenius norm we instead find $201528$(ns)].

\section{Local perturbation around the non-adiabatic transition}
\label{app:beam-splitter}

Our analysis in this appendix closely mirrors Ref.~\cite{munoz-bauza_double-slit_2019}, with some modifications. We start with the Redfield Eq.~\eqref{eq:adiabatic_tcl2} and define Liouville operators
\bes
\begin{align}
    \mathcal{L}_A  &= i\frac{t_f\Omega\pqty{s}}{2}\comm{Z}{\cdot} \label{app:H_adiabatic}\\
    \mathcal{L}_G  &= -i\frac{\dot{\theta}}{2}\comm{Y}{\cdot} \\
    \mathcal{L}_R  &= -\sum_\alpha(g_\alpha t_f)^2 \comm{S_\alpha\pqty{s}}{\Lambda_{\alpha}(s)\cdot} + \textrm{h.c.} \ ,
\end{align}
\ees
which represents the adiabatic, geometric and Redfield parts in the ME. Now we present a perturbation method for the evolution across the region $[\mu_\theta-c\alpha_\theta, \mu_\theta+c\alpha_\theta]$. First we rotate the equation with respect to the adiabatic parts of the Hamiltonian and denote the resulting Liouville operators as $\mathcal{L}_{\tilde{G}}$ and $\mathcal{L}_{\tilde{R}}$. 

Using any unitarily invariant norm, such as the operator norm, we can bound
\bes
\begin{align}
    \norm{\mathcal{L}_{\tilde{R}}\pqty{\rho}} &\leq 4\sum_\alpha (g_\alpha t_f)^2 \norm{S_\alpha(s)}\norm{\Lambda_\alpha(s)} \norm{\rho}_1 
    \label{app:redfield_superoperator_norm_step_one}\\
    & \leq \sum_\alpha (2 g_\alpha t_f \norm{S_\alpha})^2 \int_0^s \dd{s'} \abs{C_\alpha\pqty{s,s'}} 
\end{align}
\ees
where we used $\norm{\rho}_1 =1$ (trace norm), and we made use of Eq.~\eqref{eq:TCL_lambda}.
We define $g=\max_{\alpha}g_\alpha$ and note that we can always choose a normalization such that $S_\alpha\leq 1$. Then
\begin{equation}
\label{app:bound_redfield_superoperator}
    \norm{\mathcal{L}_{\tilde{R}}\pqty{\rho}} \leq (2 g t_f)^2  \int_0^s \dd{s'} \abs{C\pqty{s,s'}} \ .
\end{equation}
Noting that the correlation function is translation-invariant 
\begin{equation}
    C(s, s') = C(s-s') \ ,
\end{equation}
the bound \eqref{app:bound_redfield_superoperator} can be further simplified by using the following inequality
\begin{equation}
    \int_0^s \dd{s'} \abs{C(s, s')} < \int_0^\infty \dd{s} \abs{C(s)} \ .
\end{equation}
Defining 
\begin{equation}
    \frac{1}{\tau_{SB}}  = t_f \int_0^{\infty} \abs{C\pqty{s}} \dd{s}
\end{equation}
where $\tau_{SB}$ can be interpreted as the fastest system decoherence timescale~\cite{mozgunov_completely_2019},
the final expression becomes
\begin{equation}
    \norm{\mathcal{L}_{\tilde{R}}(\rho)} \leq 4 g^2 t_f/\tau_{SB} .
\end{equation}
This bound allows us to rigorously establish conditions under which it is safe to drop the dissipative part of the evolution. 

Next, using the bound above, we write down the evolution operator and its first order Magnus expansion:
\bes
\begin{align}
\label{eq:app_magnus_expansion-a}
    & \mathcal{E}\pqty{\mu_\theta+c\alpha_\theta,\mu_\theta-c\alpha_\theta} =  \\
\label{eq:app_magnus_expansion-b}
    & \quad \mathcal{T}\exp{-i\int_{\mu_\theta-c\alpha_\theta}^{\mu_\theta+c\alpha_\theta} \mathcal{L}_{\tilde{G}}\pqty{\tau} + \mathcal{L}_{\tilde{R}}\pqty{\tau} \dd{\tau}} =  \\
\label{eq:app_magnus_expansion-c}
    & \quad \exp{-i\int_{\mu_\theta-c\alpha_\theta}^{\mu_\theta+c\alpha_\theta} \mathcal{L}_{\tilde{G}}\pqty{\tau} \dd{\tau} + O\pqty{4 \alpha_\theta t_f g^2/ \tau_{SB}}} \ . 
    \end{align}
\ees
We made use of the formal Magnus expansion of the superoperator in going from line~\eqref{eq:app_magnus_expansion-b} to line~\eqref{eq:app_magnus_expansion-c}, wherein the lowest order term in the time-ordered propagator is the argument of line~\eqref{eq:app_magnus_expansion-b}, and the order term is a commutator of the two Liouville operators. 
As long as $\alpha_\theta t_f g^2/ \tau_{SB}\ \ll 1$, which requires either weak coupling or small $\alpha_\theta$ (as confirmed for the two examples mentioned in Appendix~\ref{app:multi_qubits_example}), we may ignore the dissipative part due to $\mathcal{L}_{\tilde{R}}$. After dropping this term we are left with
\begin{equation}
    \exp{-i\int_{\mu_\theta-c\alpha_\theta}^{\mu_\theta+c\alpha_\theta} \mathcal{L}_{\tilde{G}}\pqty{\tau} \dd{\tau}} \cdot \rho = \tilde{U} \rho \tilde{U}^\dagger \ ,
\end{equation}
where the unitary is
\begin{align}
    &\tilde{U}(\mu_\theta+c\alpha_\theta,\mu_\theta-c\alpha_\theta) = \notag\\
    &\quad\exp{-\frac{i}{2}\int_{\mu_\theta-c\alpha_\theta}^{\mu_\theta+c\alpha_\theta}\dot{\theta}(s) \tilde{Y}(s) \dd{s}} \label{app:local_unitary}\ ,
\end{align}
and $\tilde{Y}(s)$ is $Y$ in the interaction picture generated by $\mathcal{L}_A$ [Eq.~\eqref{app:H_adiabatic}]:
\begin{align}
    \tilde{Y}(s) &= U^\dagger_A(s) Y U_A(s) \notag\\
    &= -ie^{-it_f \int_0^s \Omega(s') \dd{s'}} S_+ + \text{h.c}\ ,
\end{align}
where $S_+ = (\sigma^x+i\sigma^y)/2$. Substituting $\tilde{Y}(s)$ into Eq.~\eqref{app:local_unitary}, we have
\begin{equation}\label{app:interaction_unitary}
    \tilde{U}(\mu_\theta+c\alpha_\theta,\mu_\theta-c\alpha_\theta) = \exp{-\phi S_+ + \text{h.c}}
\end{equation}
where 
\begin{equation}
    \phi = \frac{1}{2}\int_{\mu_\theta-c\alpha_\theta}^{\mu_\theta+c\alpha_\theta}\dot{\theta} e^{-it_f \int_0^s\Omega\pqty{s'}\dd{s'}} \dd{s} \ .
\end{equation}
Because $\dot{\theta}$ is highly localized within the integral limit, we can further simplify the expression as
\begin{align}
    \phi &= \frac{1}{2}\int_{-\infty}^\infty \dot{\theta} e^{-it_f \int_0^{\mu_\theta} \Omega\pqty{s'} \dd{s'}} \dd{s} =\frac{1}{2} \int_{-\infty}^\infty \dot{\theta} e^{-it_f \mu_\theta \tilde{\Omega} s} \dd{s}\ , \label{app:phi}
\end{align}
where 
\begin{equation}
    \tilde{\Omega} = \frac{\int_0^{\mu_\theta} \Omega\pqty{s'} \dd{s'}}{\mu_\theta} \ .
\end{equation}
Using Eq.~\eqref{eq:gaussian_angular_momentum}, together with the boundary condition $\theta(1) = \pi / 2$, the integral in Eq.~\eqref{app:phi} can be carried out, yielding
\begin{equation}
    \phi = \frac{\pi}{4} e^{-\pqty{t_f/t_{\text{ad}}}^2 + i\mu_\theta t_f \tilde{\Omega}} \ ,
\end{equation}
where $t_{\text{ad}} = \sqrt{2}/(\alpha_\theta \mu_\theta \tilde{\Omega})$ [Eq.~\eqref{eq:varphi}]. Finally, we can write down the matrix form of the unitary \eqref{app:interaction_unitary}:
\bes
\begin{align}
    \tilde{U} &= \exp{-\phi S_+ + \text{h.c}} \\
    &=\begin{pmatrix}
        \cos(\abs{\phi}) &
        - \sin(\abs{\phi})e^{i\mu_\theta t_f \tilde{\Omega}} \\
         \sin(\abs{\phi})e^{-i\mu_\theta t_f \tilde{\Omega}} &
        \cos(\abs{\phi}) 
        \end{pmatrix} \\ 
    &=
    U_A^\dagger\pqty{\mu_\theta}\begin{pmatrix}
    \cos(\abs{\phi}) &
    - \sin(\abs{\phi}) \\
     \sin(\abs{\phi}) &
    \cos(\abs{\phi}) 
    \end{pmatrix} U_A\pqty{\mu_\theta}\ . 
\end{align}
\ees
If we rotate the unitary back into the adiabatic frame, it becomes Eq.~\eqref{eq:U}.

As a final remark, in the limit of $\alpha_\theta \to 0$, the unitary non-adiabatic transition has the same effects as an ideal beam splitter:
\begin{equation}
    U \to \frac{\sqrt{2}}{2}(I-iY) \ .
\end{equation}

\section{Derivation of Eq.~\eqref{eq:app_dFg}}
\label{app:dFg}

Taking the partial derivative of Eq.~\eqref{eq:F_g} with respect to $s_p$, we have
\bes
\begin{align}
    \pdv{\mathcal{F}_g}{s_p} &= \int_{\mu^+_\theta}^{\tau_f}\dd{\tau} \Bqty{\pdv{\Gamma}{s_p}G + \Gamma \pdv{G}{s_p}} \\
    &= \int_{\mu^+_\theta}^{\tau_f}\dd{\tau} G \Bqty{\pdv{\Gamma}{s_p}-\Gamma\int_{\tau}^{\tau_f} \pdv{X}{s_p}\dd{\tau'}} \ . \label{eq:pdv_fg}
\end{align}
\ees
Denoting either $\Gamma(\tau)$ or $X(\tau)$ by $F(\tau)$, their partial derivatives with respect to $s_p$ are given by (for a proof see Appendix~\ref{app:proof49}): 
\begin{equation}
\label{eq:app_derivative}
    \pdv{F(\tau)}{s_p} = \begin{cases}
    0 \quad &\tau \leq s_p \\
    \bar{F}'(s_p) \quad &s_p < \tau \leq s_p + s_d \\
    0 & s_p + s_d < \tau \leq 1+s_d
    \end{cases}
\end{equation}
where
    $\bar{F}'(s_p) = \dv{\bar{F}(s)}{s}\big\vert_{s=s_p}$.
This can also be thought as the chain rule
\begin{equation}
    \pdv{F(\tau)}{s_p} = \dv{\bar{F}(s)}{s} \pdv{s(\tau)}{s_p} \ ,
\end{equation}
where $\pdv{s(\tau)}{s_p}$ follows by differentiating Eq.~\eqref{eq:cases}
\begin{equation}
    \pdv{s(\tau)}{s_p} = \begin{cases}
    0\quad &\tau \leq s_p \\
    1 \quad &s_p < \tau \leq s_p + s_d \\
    0 \quad &s_p + s_d < \tau \leq 1+s_d
    \end{cases} .
    \label{eq:Dcases}
\end{equation}
Let us consider the two end points of Eq.~\eqref{eq:pdv_fg}, $\pdv{\mathcal{F}_g}{s_p} \big\vert_{s_p=\mu_\theta^+}$ and $\pdv{\mathcal{F}_g}{s_p} \big\vert_{s_p=1}$. 

\subsection{Derivation of Eq.~\eqref{eq:app_dFg_x}} 
Consider $\pdv{\mathcal{F}_g}{s_p} \big\vert_{s_p=\mu_\theta^+}$. This integral can be simplified by realizing that both $\pdv{\Gamma}{s_p}\big\vert_{s_p=\mu_\theta^+}\pqty{\tau}$ and $\pdv{X}{s_p}\big\vert_{s_p=\mu_\theta^+}\pqty{\tau}$ are rectangular functions within the pausing region $\bqty{\mu^+_\theta, \mu^+_\theta+s_d}$ [Eq.~\eqref{eq:app_derivative}]. As a result, $\pdv{\Gamma}{s_p}\big\vert_{\mu_\theta^+}(\tau)$ and $\int_{\tau}^{\tau_f}\pdv{X}{s_p}\big\vert_{\mu_\theta^+}(\tau')\dd{\tau'}$ are zero when $\tau > \mu^+_\theta+s_d$. The former follows directly from the definition and the latter is because the integrand is zero in the entire region of integration. This allows us to change the integration limit from $\tau_f$ to $\mu_\theta^+ + s_d$ in equation \eqref{eq:pdv_fg} and write
\begin{align}
  &  \pdv{\mathcal{F}_g}{s_p} \Big\vert_{s_p=\mu_\theta^+} = \notag \\
   & \quad  \int_{\mu^+_\theta}^{\mu_\theta^++s_d}\dd{\tau} G(\tau) \bigg(\pdv{\Gamma}{s_p}\Big\vert_{\mu_\theta^+}-\Gamma\int_{\tau}^{\tau_f} \pdv{X}{s_p}\Big\vert_{\mu_\theta^+}\dd{\tau'}\bigg) = \nonumber \\
    &\quad \int_{\mu^+_\theta}^{\mu_\theta^++s_d}\dd{\tau} G(\tau) \bigg(\bar{\Gamma}'\pqty{\mu_\theta^+} 
    \notag \\
    &\qquad 
    -\bar{\Gamma}\pqty{\mu_\theta^+} \bar{X}'(\mu_\theta^+)(s_d+\mu^+_\theta-\tau)\bigg) \ .     \label{eq:app_dFg_1}
\end{align}
To obtain the second equality above, we explicitly carried out the integration
\begin{align}
    \int_\tau^{\tau_f}\pdv{X\pqty{\tau'}}{s_p}\Big\vert_{s_p=\mu_\theta^+} \dd{\tau'} &= \bar{X}'(\mu_\theta^+) \int_\tau^{\mu_\theta^+ + s_d} 1 \dd{\tau'} \notag\\
    &=\bar{X}'(\mu_\theta^+)(s_d +\mu^+_\theta - \tau) \ .
    \label{eq:52}
\end{align}
Using Eq.~\eqref{eq:G} and noticing that $X(\tau')$ is constant for $\tau'\in[\mu_\theta^+,\tau]$ with $\tau \in [\mu_\theta^+, \mu_\theta^++s_d]$ (the pausing region), we have:
\begin{align}
    G(\tau)\big\vert_{s_p=\mu_\theta^+} &= \exp{\pqty{\int_{\mu_\theta^+}^{\tau} -\int_{\mu_\theta^+}^{1+s_d}}X(\tau')\dd{\tau'}} \notag \\
    &=\exp{\bar{X}\pqty{\mu_\theta^+}\pqty{\tau-\mu_\theta^+}}G(\mu_\theta^+) .
\end{align}
Thus Eq.~\eqref{eq:app_dFg_1} can be further simplified with a change of variable $\tau=s_d x + \mu_\theta^+$, upon which we arrive at Eq.~\eqref{eq:app_dFg_x}:
\begin{align}
    &  \pdv{\mathcal{F}_g}{s_p} \Big\vert_{s_p=\mu_\theta^+} =  \notag \\
      &\quad s_d G\pqty{\mu_\theta^+}\int_{0}^{1}\dd{x} e^{\bar{S}x} \bigg(\bar{\Gamma}'\pqty{\mu_\theta^+}  
      -s_d\bar{\Gamma}\pqty{\mu_\theta^+} \bar{X}'(\mu_\theta^+)(1-x)\bigg) \notag\ ,     
  \end{align}
  where $\bar{S}=s_d\bar{X}\pqty{\mu_\theta^+}$.

\subsection{Derivation of Eq.~\eqref{eq:app_dFg_4}} 
Consider $\pdv{\mathcal{F}_g}{s_p} \big\vert_{s_p=1}$. The expression at this end point can similarly be obtained by splitting the integral into two parts $\int_{\mu_\theta^+}^{\tau_f}d\tau = \int_{\mu_\theta^+}^{1} d\tau+ \int_{1}^{1+s_d}d\tau$ [recall, per Eq.~\eqref{eq:tau_f}, that $\tau_f = 1+s_d$]. The integrands of the first integral $\int_{\mu_\theta^+}^{1}d\tau$ satisfy:
\bes
\begin{align}
\label{eq:53a}
    \pdv{\Gamma\pqty{\tau}}{s_p}\Big\vert_{s_p=1} &= 0 \\
\label{eq:53b}
    \int_{\tau}^{\tau_f} \pdv{X\pqty{\tau'}}{s_p}\Big\vert_{s_p=1} \dd{\tau'} &= \bar{X}'(1) \int_{1}^{\tau_f} 1 \dd{\tau'} \notag \\
    & = s_d \bar{X}'(1) \ .
\end{align}
\ees
Eq.~\eqref{eq:53b} follows from the same reasoning as Eq.~\eqref{eq:52}.
Eq.~\eqref{eq:53a} follows from the chain rule applied to Eq.~\eqref{eq:Gammatau}, which gives $\pdv{s(\tau)}{s_p}\Big\vert_{s_p=1} = 0$ (Eq.~\eqref{eq:Dcases}) as an overall prefactor since the integration over $\tau$ goes up to $\tau=1$.

Again using Eqs.~\eqref{eq:52} and~\eqref{eq:Dcases}, the integrands of the second integral $\int_{1}^{1+s_d}d\tau$ satisfy: 
\bes
\begin{align}
    \pdv{\Gamma\pqty{\tau}}{s_p}\Big\vert_{s_p=1} &= \bar{\Gamma}'(1) \\
    \int_\tau^{\tau_f}\pdv{X\pqty{\tau'} }{s_p}\Big\vert_{s_p=1}\dd{\tau'} &= \bar{X}'(1) \int_\tau^{\tau_f} 1 \dd{\tau'} \notag\\
    &=\bar{X}'(1)(s_d +1 - \tau) \ .
\end{align}
\ees
Combining these results into Eq.~\eqref{eq:pdv_fg}, the final expression becomes Eq.~\eqref{eq:app_dFg_4}:
\begin{align}
&    \pdv{\mathcal{F}_g}{s_p} \bigg\vert_{s_p=1} = -s_d \bar{X}'\pqty{1}\int_{\mu^+_\theta}^{1}\dd{\tau}G\pqty{\tau}\Gamma\pqty{\tau} \notag \\
    &\quad + \int_{1}^{1+s_d}\dd{\tau}G\pqty{\tau}\Big(\bar{\Gamma}'(1) -\bar{\Gamma}(1)\bar{X}'(1)\pqty{s_d+1-\tau} \Big) \notag\ .
\end{align}

\section{Proof of Eq.~\eqref{eq:app_derivative}}
\label{app:proof49}

\begin{figure}[htb]
\includegraphics[width=\columnwidth]{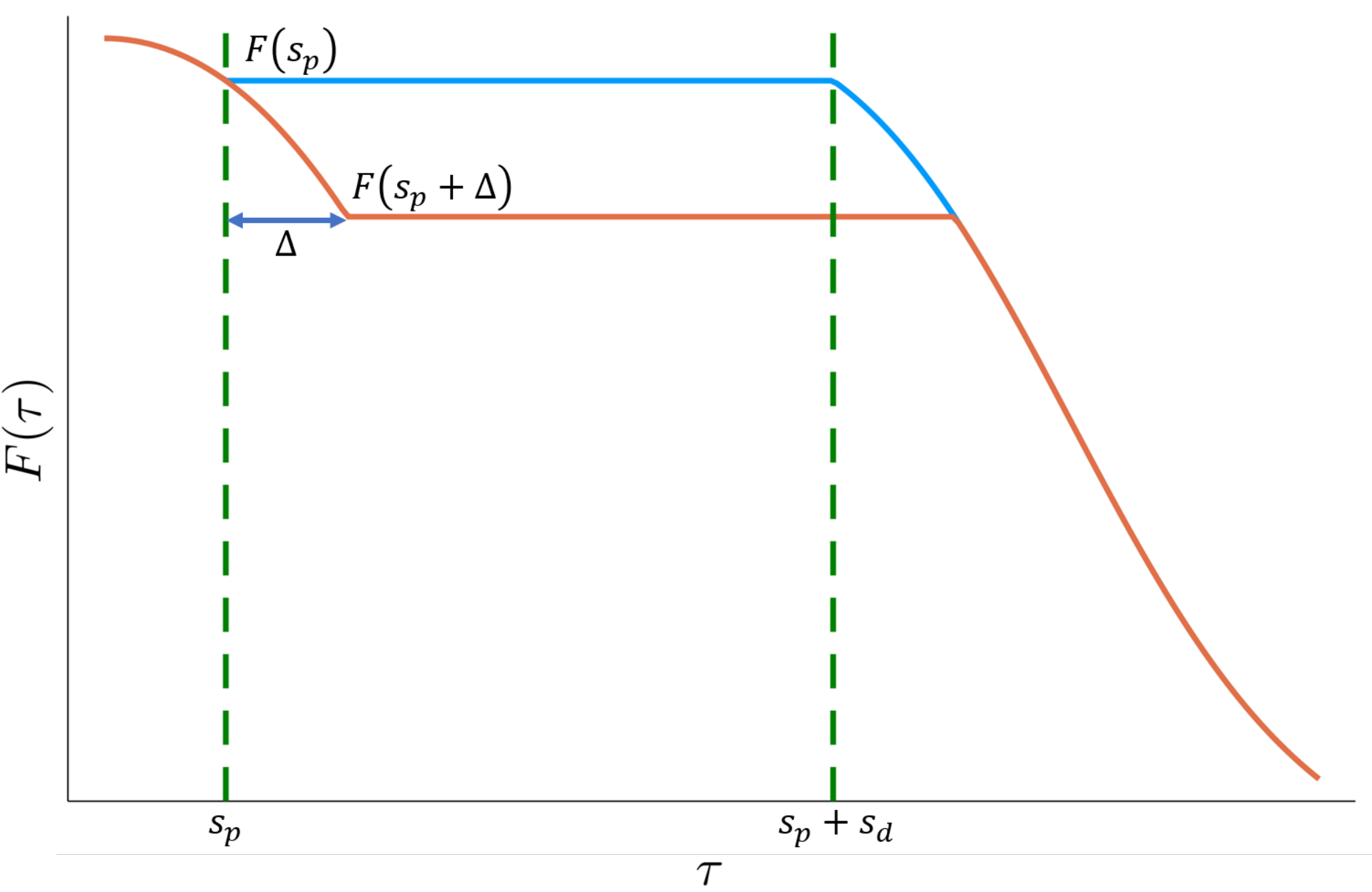}
\caption{Graphical proof of Eq.~\eqref{eq:app_derivative}. The partial derivative is obtained in the limit of $\Delta \to 0$.
}
\label{fig:partial_derivative}
\end{figure}

To prove Eq.~\eqref{eq:app_derivative}, we start from the definition of the partial derivative:
\begin{equation}
    \pdv{F}{s_p} = \lim_{\Delta \to 0}\frac{F\pqty{\tau, s_p+\Delta, s_d} - F\pqty{\tau, s_p, s_d}}{\Delta} \ .
\end{equation}
From Fig.~\ref{fig:partial_derivative}, we see that in the limit $\Delta \to 0$
\begin{equation}
    F(\tau, s_p+\Delta, s_d) = \begin{cases}
    F\pqty{s\pqty{\tau}} \quad &\tau < s_p \\
    F(s_p+\Delta) \quad &s_p < \tau < s_p + s_d \\
    F\pqty{s\pqty{\tau}} & s_p + s_d < \tau < 1+s_d
    \end{cases} \ , 
\end{equation}
so
\begin{equation}
    \pdv{F}{s_p} = \lim_{\Delta \to 0}\frac{F\pqty{s_p+\Delta} - F\pqty{s_p}}{\Delta} = F'(s=s_p)
\end{equation}
for $s_p < \tau < s_p + s_d$ and zero elsewhere.

\section{Proof of Eq.~\eqref{eq:barSbound}}
\label{app:barSbound}

To prove Eq.~\eqref{eq:barSbound} we use
\begin{equation}
    f(x) = x + W_{-1}(-e^{-1/x}/x) > 0 \quad \forall x>1 \ ,
\end{equation}
which can be proved with $f(1)=0$ and
\bes
\begin{align}
\label{eq:lambert_derivative}
    f'(x) &= 1 + \frac{W_{-1}(-e^{-1/x}/x)}{1+W_{-1}(-e^{-1/x}/x)}\frac{1-x}{x^2} \\
 \label{eq:lambert_derivative2}
   &= \frac{\abs{1+W_{-1}}x^2 - |W_{-1}|x + |W_{-1}|}{\abs{1+W_{-1}}x^2} \\
\label{eq:lambert_derivative3}
   & > 0 \quad \forall x>1 \ .
\end{align}
\ees
Line~\eqref{eq:lambert_derivative} is obtained using the derivative formula
\begin{equation}
    \dv{W(z)}{z} = \frac{W(z)}{z\pqty{1+W(z)}}
\end{equation}
and line~\eqref{eq:lambert_derivative2} comes from Eq.~\eqref{eq:W_minus_branch}. To prove line~\eqref{eq:lambert_derivative3}, we consider the quadratic function in the numerator of line~\eqref{eq:lambert_derivative2}, whose roots are:
\begin{equation}\label{eq:roots}
    x_{\pm} = -\frac{\abs{W_{-1}}\pm\sqrt{-3\abs{W_{-1}}^2+4\abs{W_{-1}}}}{2\abs{1+W_{-1}}} \ .
\end{equation}
For $\abs{W_{-1}}>4/3$, the discriminant is smaller than zero. For $1<\abs{W_{-1}}\leq4/3$, the discriminant lies within $[0, 1)$ so both roots are negative. In both cases, the quadratic function is positive for $x>1$, which leads to line~\eqref{eq:lambert_derivative3}.

\section{Demonstration that $\epsilon \ll 1$}
\label{app:proof_epsilon}

To show that $\epsilon \ll 1$ for $\epsilon \equiv \frac{\bar{\Gamma}\pqty{1}\beta \bar{\Omega}'(1)}{|\bar{\Gamma}'(1)|(1+e^{\beta\bar{\Omega}(1)})}$ [Eq.~\eqref{eq:epsilon}], we start with the expression for $\bar{\Gamma}\pqty{s}$ [Eq.~\eqref{eq:Gammatau}]. Because $\sum_\alpha \abs{S_\alpha^{01}(s)}^2 = 1$ for the system-bath coupling operators in our model Eq.~\eqref{eq:constructed_coupling}, we obtain 
\begin{equation}
    \bar{\Gamma}\pqty{s}=\bar{\Gamma}[\bar{\Omega}\pqty{s}]=(1+s_d)t_f\gamma[\bar{\Omega}\pqty{s}] \ ,
\end{equation}
where we have made the gap dependence of $\bar{\Gamma}\pqty{s}$ explicit.

Next, $\epsilon$ can be simplified as
\bes
\begin{align}
    \epsilon &= \frac{\bar{\Omega}'(1)\bar{\Gamma}(1)}{\abs{\bar{\Gamma}'(1)}} \frac{\beta}{1+e^{\beta\bar{\Omega}\pqty{1}}}\\
    &= \frac{\gamma[\bar{\Omega}\pqty{1}]}{|\dv{\gamma}{\bar{\Omega}}[\bar{\Omega}\pqty{1}]|}\frac{\beta}{1+e^{\beta\bar{\Omega}\pqty{1}}}\ , 
    \label{eq:epsilon_1}
\end{align}
\ees
where we used the chain rule to write $\bar{\Gamma}'\pqty{s} = (1+s_d)t_f \bar{\Omega}'\pqty{s} \dv{\gamma}{\bar{\Omega}}$.

Any spectral density with an exponential high-frequency cutoff $e^{-\omega/\omega_c}$, such as the Ohmic bath  [Eq.~\eqref{eq:Ohmic}], will result in 
\beq
\label{eq:ohmic_constant}
\frac{\gamma[\bar{\Omega}\pqty{1}]}{|\dv{\gamma}{\bar{\Omega}}[\bar{\Omega}\pqty{1}]|} \sim \omega_c \quad \text{for } \omega > \omega_c \ .
\eeq
Thus, as long as $\bar{\Omega}\pqty{1} > \omega_c,1/\beta$ we find that $\epsilon$ is exponentially small in the final energy gap $\bar{\Omega}\pqty{1}$.

\begin{figure}[b]
    \includegraphics[width=\columnwidth]{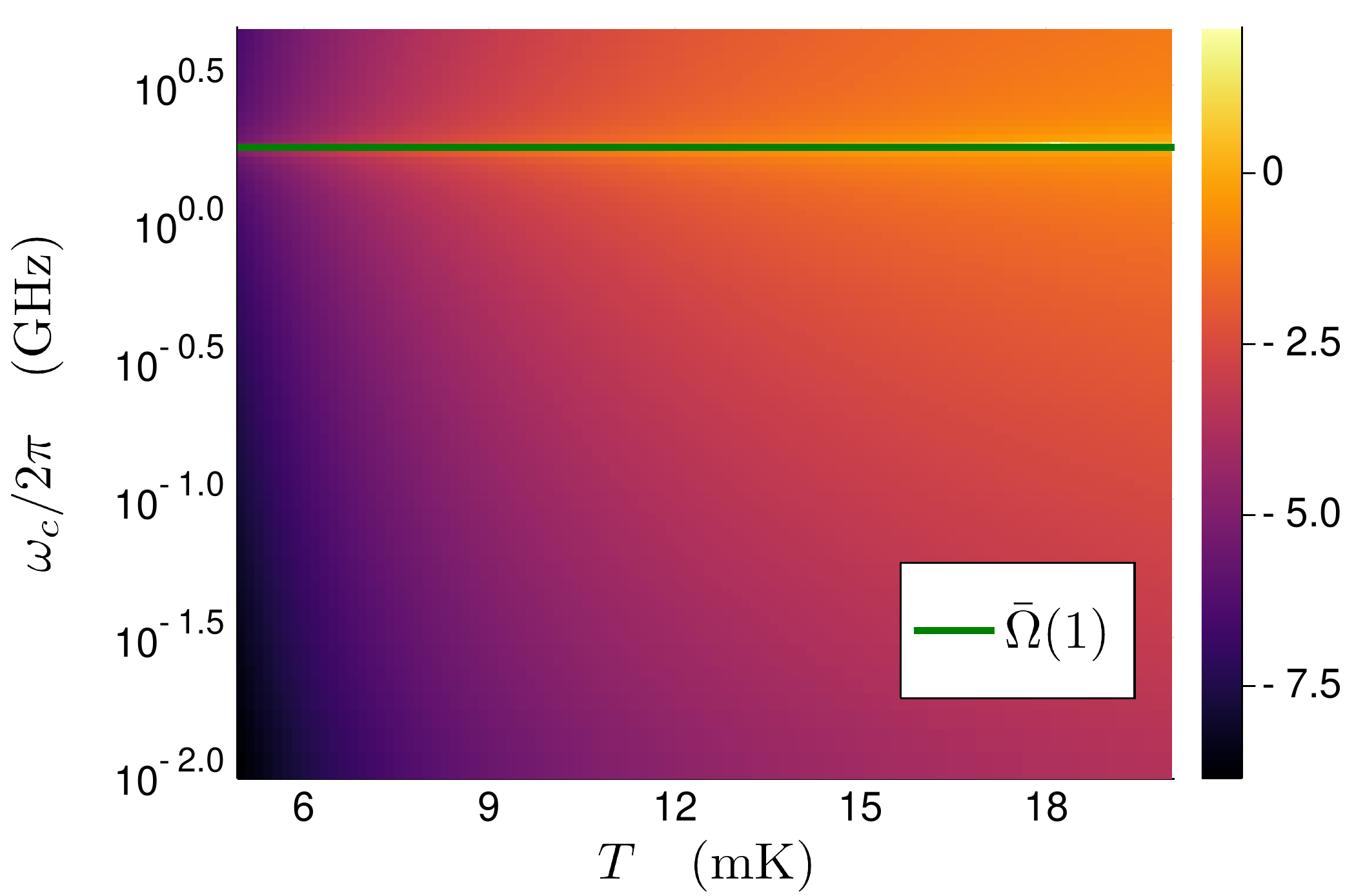}
    \caption{Heat map of $\log_{10}\epsilon$ for different $\omega_c$ and $T$ values for the Ohmic bath spectrum [Eq.~\eqref{eq:Ohmic}], covering the entire parameter region shown in Fig.~\ref{fig:Gamma_s}. The gap $\bar{\Omega}\pqty{1}\approx 1.88(\mathrm{GHz})$ is chosen in accordance with parameters used in Fig.~\ref{fig:schedules}. The maximum of $\epsilon$ appears near the line $\omega_c = 2\pi\bar{\Omega}\pqty{1}$.
    }
    \label{fig:epsilon}
\end{figure}

We plot $\log_{10}\epsilon$ for different temperatures and cutoff frequencies for the Ohmic case in Fig.~\ref{fig:epsilon}, and confirm that for reasonable parameters indeed $\epsilon \ll 1$.
    
This argument fails if $\bar{\Gamma}'(s^*) = 0$ for $s^*\in [\mu_\theta^+,1]$. For such cases, we only need to shift the end point from $1$ to $s^*$. Then the optimal pausing position will be in the interval $[\mu_\theta^+, s^*]$.

\bibliography{refs}
\end{document}